\documentstyle[aps,preprint,epsfig,amssymb]{revtex}  

\newcommand{\be}{\begin{equation}} 
\newcommand{\ee}{\end{equation}} 
\newcommand{\bea}{\begin{eqnarray}} 
\newcommand{\eea}{\end{eqnarray}} 
\newcommand{\bfig}{\begin{figure}} 
\newcommand{\efig}{\end{figure}} 
\begin{document}       
  
\title{Front propagation in laminar flows}

\author{M. Abel$^{a,d}$, A.~Celani$^{b}$, D. Vergni $^{a,c}$  and A. Vulpiani $^{a,c}$}

\address{$^a$ Dipartimento di Fisica, Universit\'a di Roma ''La Sapienza''\\
Piazzale Aldo Moro 2, I-00185 Roma, Italy}
\address{$^b$ CNRS, INLN, 1361 Route des Lucioles ,
F-06560 Valbonne, France.}
\address{$^c$ INFM unit\`a di Roma ''La Sapienza''
Piazzale Aldo Moro 2, I-00185 Roma, Italy}
\address{$^d$ University of Potsdam, 14469 Potsdam, Germany}

\maketitle

\begin{abstract}

The problem of front propagation in flowing media is addressed for
laminar velocity fields in two dimensions. 
Three representative cases are discussed:
stationary cellular flow, stationary shear flow, and percolating
flow. Production terms of Fisher-Kolmogorov-Petrovskii-Piskunov type
and of Arrhenius type are considered under the assumption of no 
feedback of the concentration on the velocity. Numerical simulations of
advection-reaction-diffusion equations have been performed by an
algorithm based on discrete-time maps. The results show
a generic enhancement of the speed of front propagation by
the underlying flow. For small molecular diffusivity,
the front speed $V_f$ depends on the typical
flow velocity $U$ as a power law with an exponent depending
on the topological properties of the flow, and on the ratio of
reactive and advective time-scales. 
For open-streamline flows we find 
always $V_f \sim U$, whereas for cellular flows
we observe $V_f \sim U^{1/4}$ for fast advection, and
$V_f \sim U^{3/4}$ for slow advection.

\vspace{0.5cm}
\end{abstract}
\newpage
\section{Introduction}
\label{sec:1}

Interface motion and front propagation in fluids occur in many
different areas of interest to science and technology.
Among the most important examples we mention chemical reaction
fronts in liquids, population dynamics of ecological communities
(e.g. plankton in the ocean), atmospheric chemistry (ozone hole)
and flame propagation in gases~\cite{flame}.\\
The mathematical description of those
phenomena is based on partial
differential equations (PDE) for the evolution of the concentration
of the reacting species and the evolution of the velocity field~\cite{combus}.
In principle the two PDEs (for the reactants and the velocity field)
are coupled, often in a nontrivial way.
An example is given by a recent study of reactants coupled to the Navier-Stokes
equation by a Boussinesq term~\cite{Malham}.
A mathematical simplification can be obtained assuming that the
reactants do not influence the velocity field which evolves independently.
In such a limit the dynamics is still completely nontrivial and it is
described by a so-called advection-reaction-diffusion equation. 
In the most compact model one considers a single scalar field 
$\theta({\mathbf x},t)$, which represents the fractional concentration
of products. The field $\theta$ has a zero value 
in the regions containing fresh material only, 
$\theta$ is unity where the reaction is over and there are only
inert products left. 
In the region where the production takes place and reactants and products
co-exist, the field $\theta$ 
assumes intermediate values.\\
The evolution of $\theta$ in a reacting fluid with molecular 
diffusivity $D_0$ is described by the PDE
\be
   \frac{\partial}{\partial t} \theta + 
   ({\mathbf u} \cdot {\mathbf \nabla}) \theta =
   D_0 \nabla^2 \theta + \frac{1}{\tau_r} f(\theta)
   \label{eq:rad}
\ee
where ${\mathbf u}({\mathbf x},t)$ is an incompressible velocity field.
The second term on the r.h.s. of Eq.~(\ref{eq:rad}) describes
the production process, characterized by a typical time $\tau_r$.
The shape of $f(\theta)$ depends on the phenomenon under
investigation, and we will consider two relevant functional forms:
\begin{itemize}
   \item [1)] $f(\theta) = \theta(1 - \theta)$, or more generally
	any convex function ($f''(\theta) < 0$) such that $f(0)=f(1)=0$, 
	$f'(0) > 0$ and $f'(1) < 0$. This is called the 
	Fisher-Kolmogorov-Petrovskii-Piskunov 
        (FKPP) nonlinearity~\cite{kpp}.
	The production term is proportional to the concentration of 
	reactants, $1-\theta$, and to the concentration of products, $\theta$.
   \item [2)] $f(\theta) = e^{\scriptsize {-\theta_c / \theta}}(1 - \theta)$.
              This is the Arrhenius nonlinearity~\cite{xin2000}.
              In this case, the solution $\theta = 0$ is only 
	      marginally unstable,
              and the parameter $\theta_c$ plays the role
              of an activation concentration, since practically
 	      no production takes place for products concentrations below
	      that threshold. Production is still proportional
	      to the concentration of fresh material.
\end{itemize}
We will always consider initial conditions such that
$\theta({\mathbf x},0)\to 1$ exponentially fast as $x$, the horizontal 
coordinate, approaches $-\infty$, and $\theta({\mathbf x},0)\to 0$
exponentially fast as $x \to \infty$. The initial profile of $\theta$
has no variation along the transversal axis. 
The choice of such an initial condition, for
which the concentration of products has a noncompact support, is of
interest because it suppresses all possible flame-quenching effects
that may appear in case $2)$ (see e.g. \cite{quenching}).
This allows a direct comparison between FKPP and Arrhenius
production terms. \\
There exists a huge literature for the case ${\mathbf u}=0$ 
(see e.g. \cite{xin2000}). In that case the physical mechanism 
for front propagation resides in the
the combined effect of diffusion and production.  
Let us indeed consider a one-dimensional situation 
where a reservoir of fresh material is located on the right side,
whereas on the opposite side we have only inert products. At the boundary
between the two phases, diffusion mixes 
fresh material and inert products, broadening the interface. 
Then, production raises the level of the concentration of products, thus
shifting the interface, to the right in this case.
The final result is a front propagating from left to right,
eating out fresh material to leave behind digested, inert products. 
The front speed at large times reaches
a limiting value $V_0$. For the FKPP nonlinearity one has the exact result
\be
V_0 = 2 \sqrt{\frac{D_0 f'(0)}{\tau_r}}
\label{eq:v0}
\ee
whereas for a generic $f(\theta)$ one has 
the bounds~\cite{kpp,xin2000,AronWein78}
\be
2 \sqrt{\frac{D_0}{\tau_r} f'(0)} \leq V_0 \leq 
2 \sqrt{\frac{D_0}{\tau_r} \sup_\theta \frac{f(\theta)}{\theta}}\,\,.
\label{eq:boundv0}
\ee
It has to be remarked that the convergence to the 
limiting velocity is extremely slow for FKPP production 
\cite{Armero97,vansa}, therefore this case requires special attention
especially for nonuniform flow.
\\
In the presence of a moving medium, i.e. ${\mathbf u} \neq 0$, one expects
that the front propagates with an average limiting speed $V_f$.  A
problem of primary interest is to determine the dependence of $V_f$ on
the properties of the velocity field ${\mathbf u}$~\cite{CW,KA}.
In this article we consider front propagation
in simple laminar flows (shear flow and systems with cellular structures)
which, in spite of their apparent simplicity, show intriguing 
behaviour~\cite{Ros,Shr,Pom,Rhi}.
For a given structure of the flow field, we aim to explore 
the dependence of $V_f$ on relevant parameters, such as 
the typical flow velocity $U$ and the production time-scale $\tau_r$. 
In terms of adimensional quantities, we look for an expression for the
speed enhancement $V_f/V_0$ in terms of the {\em Damk\"ohler number},
$Da=L/(U\tau_r)$, which measures the ratio of advective to reactive
time-scales, and in terms of the {\em P\'eclet number}, $Pe=UL/D_0$
which expresses the relative weight of advection and diffusion. We
will mainly be interested in the case of large $Pe$, to highlight
the combined effects of advection and reaction.  
The two regimes $Da \ll 1$ and $Da \gg 1$ 
are quite different in nature: in the first case, typical
for slow reaction rates or fast advection, the front interface is
distributed over several length-scales $L$, and for this reason it
goes under the name of ``distributed reaction zone'' regime
\cite{Ronney95}; in the second case, the front is thin compared to
$L$, and it propagates according to a Huyghens-like principle, hence
the name of ``geometrical optics'' regime \cite{Ronney95}.
We will provide a detailed analysis
of these two regimes, highlighting their differences. 
In the context of the thin front regime we can mention, among the many
contributions, the works on the G-equation approximation and its
relation with the ``geometrical optics'' regime~\cite{mclaughlin,embid}, 
the work on turbulent flows~\cite{CW,KA} and the numerical study 
of front propagation in synthetic turbulence~\cite{marti}.\\
A  hint to the effect of an underlying flow on front propagation is given
by the observation that the front speed increases with the square-root of
molecular diffusivity. It is well-known that diffusive transport is 
always enhanced by incompressible flow, resulting in an
effective diffusion coefficient $D_{\mbox{\scriptsize eff}}>D_0$ 
\cite{AM,AV,K70}. Therefore, 
it is reasonable to expect
that the front speed will be enhanced, too. This physical argument
can be upgraded to a mathematically rigorous statement in the case
of a ``slow'' reaction, that is for $Da \ll 1$ (see Section \ref{sec:2}).

From the mathematical viewpoint, there exist lower bounds to the
speed of the front which confirm the expectation that the flow enhances
front propagation \cite{Const,KR00}. These bounds take different
forms according to the topological structure of the flow field.
One can distinguish two main classes: cellular flows, characterized by having
closed streamlines only, and percolating flows, which possess open streamlines 
(shear flows are a particular case of this second class, having only
open streamlines). 
For cellular flows, it has been shown that $V_f/V_0 \ge C_1 Da^{-1/2}+C_2$
for $Da \ge 1$ and $V_f/V_0 \ge C_1 Da^{-1/5}+C_2$ for $Da \le 1$, 
with constants $C_1,C_2$ depending on the shape of the production term.
For percolating flows, the lower bound
is expressed as $V_f \ge K_1 U$ where again $K_1$ is a constant depending
on $f$.  The most general upper bound, valid for flows of both classes, 
is $V_f \le V_0 + K_2 U$ \cite{Const}. We therefore see that percolating 
flows are constrained to a linear dependence on the stirring intensity $U$ 
(see Section~\ref{sec:4a} for numerical results).
Cellular flows have more intriguing properties,
as we will see in Section~\ref{sec:4b}: for fast advection,
$Da \ll 1$, the front speed depends
on the flow velocity as $V_f \propto U^{1/4}$, 
to  be compared with the lower bound prediction $\propto U^{1/5}$ 
and the upper bound $\propto U$; for slow advection, $Da \gg 1$,
we obtain $V_f \propto U^{3/4}$ to compare with the lower bound 
$\propto U^{1/2}$ and the upper bound $\propto U$. 
It is clear that these bounds do not provide a sharp evaluation of
the front speed in the case of a cellular flow.

To close the overview of mathematical results, we anticipate that
a different upper bound for the front speed enhancement can be obtained by
reformulating the solutions of Eq.~(\ref{eq:rad}) in terms of a path-integral
(see Section~\ref{sec:2} for details, and references therein).
This bound yields an expression similar to the one obtained in 
the absence of any flow, but with an effective diffusion coefficient 
$D_{\mbox{\scriptsize eff}}$
replacing the molecular one. Explicitly, we show that 
\be
V_f \le 2\sqrt{{D_{\mbox{\scriptsize  eff}}\over \tau_r} 
\sup_{\theta} {f(\theta)\over\theta}}
\label{eq:upb}
\ee
where the dependence of $D_{\mbox{\scriptsize eff}}$ on the flow
parameters and on the molecular diffusion 
can be derived by the analysis of Eq.~(\ref{eq:rad}) when
the production term has been switched off. For a cellular flow,
we have the result $D_{\mbox{\scriptsize eff}}\sim \sqrt{ULD_0}$, 
\cite{Pom,Shr}, whereas for a shear flow 
$D_{\mbox{\scriptsize eff}} \sim (UL)^2/D_0$ \cite{Zeldovich}.
Inserting these latter expressions in Eq.~(\ref{eq:upb}) we obtain the
behaviour $V_f \propto U^{1/4}$ for the cellular flow 
$--$ also obtained by~\cite{Pomeau} $--$  
and $V_f \propto U$ for the shear flow, 
remarkably close to the observed ones for $Da \ll 1$. Furthermore,
the upper bound~(\ref{eq:upb}) is sharp in the regime where
 homogenization techniques apply 
(see Section~\ref{sec:2} and Appendix~\ref{app:c}). 
In other words, for fast advection
the effect of the underlying flow can be compactly expressed
in the renormalization of the diffusion coefficient.
On the contrary, for slow advection in a cellular flow, 
the front speed departs significantly 
from the upper bound, with an
increase in front speed less prominent than in the fast advection regime. 
We will see in Section~\ref{sec:4b} that 
the main physical mechanism accounting for this depletion is 
the appearance of an effective reaction term as a consequence
of the joint effect of advection and reaction.\\   
These observations lead us to argue that the effect of a
stirring velocity on front propagation can be in general summarized
in the renormalization of the two relevant parameters: {\it (i)}
an effective diffusivity, which is always larger than the molecular one 
for an incompressible flow, and {\it (ii)} 
an effective production term, which is slower than the microscopic one
for slow advection. 
\\  
The remainder of the paper is organized as follows: in Section~\ref{sec:2},
we derive some general results valid for {\em all} advecting flows;
Section~\ref{sec:3} presents the algorithm employed for the
numerical solution of Equation~(\ref{eq:rad}); 
in Section~\ref{sec:4} we discuss the results of numerical simulations;
Section~\ref{sec:5} is devoted to conclusions and discussion. Technical
and numerical details are treated in the Appendices.
\section{Upper bounds to front speed}
\label{sec:2}
In this Section we show how to establish the  upper bound~(\ref{eq:upb}) 
for the speed of front propagation in a generic incompressible flow
and a generic production term. This result is the consequence 
of the deeply rooted link existing between front propagation and 
advective transport. In other words, we will exploit the
relationship of the solutions of Eq.~(\ref{eq:rad}) to the solutions of
the same equations in absence of production terms. This will yield the
 constraint~(\ref{eq:upb}) involving the front speed $V_{f}$,
the effective diffusion coefficient $D_{\mbox{\scriptsize eff}}$
and the production time-scale $\tau_r$. 
In general this relation is 
an inequality, and not a sharp functional relation. In a graphical form, 
this amounts to say that
``front propagation$\neq$(advection$+$diffusion)$+$production''.
A significant exception to this general rule is given by the limit of very 
slow reaction (or very fast advection), where the bound~(\ref{eq:upb})
becomes sharp. In this case, homogenization techniques, detailed
in Appendix~\ref{app:a}, allow to show that the problem of the
determination of the front speed reduces to the problem of determining
$D_{\mbox{\scriptsize eff}}$. This is essentially due to the large separation
of typical time-scales.    

We start our proof of the inequality~(\ref{eq:upb}) by recalling the
fundamental relation among the solution of the PDE (\ref{eq:rad})
and the trajectories of particles advected by a velocity field 
${\mathbf u}({\mathbf x},t)$ and subject 
to molecular diffusion\cite{freid,fedotov}
\be
   \theta({\mathbf x}, t) = \left \langle { 
   \theta({\mathbf r}(0), 0)\, e^{\scriptsize {\,\int_0^t c(\theta({\mathbf r}(s),s)) 
   {\mathrm d}s}}} \right \rangle_\eta
   \label{eq:feykacext}
\ee
where $$c(\theta) = \frac{1}{\tau_r}\frac{f(\theta)}{\theta}$$
is the growth rate of $\theta$. The average is performed 
over the trajectories
evolving according to the Langevin equation
\be
     {d {\mathbf r}(t) \over dt}={\mathbf v}({\mathbf r}(t),t)+\sqrt{2D_0}
	 \mbox{\boldmath $\eta$}(t)
\label{eq:langevin}
\ee
with final conditions  ${\mathbf r}(t) = {\mathbf x}$. The white noise
term $\sqrt{2D_0}\mbox{\boldmath $\eta$}(t)$ 
accounts for molecular diffusion.\\
Since the growth rate is always bounded from above, 
$c(\theta) \le c_{\mbox{\scriptsize  max}}\equiv\sup_\theta c(\theta) $, 
Eq.~(\ref{eq:feykacext})
yields the inequality 
\be
\theta(t,{\mathbf x})\le 
\left\langle \theta(0,{\mathbf r}(0) \right\rangle 
\exp(c_{\mbox{\scriptsize max}}t)\;.
\label{eq:ineq}
\ee
For FKPP production terms the maximum occurs for $\theta=0$, 
that is we have $c(\theta) \le c(0)=f'(0)/\tau_r=1/\tau_r$, 
therefore $c_{\mbox{\scriptsize max}}=1/\tau_r$.
In the inequality~(\ref{eq:ineq}), the term in angular brackets denotes the 
probability that the trajectory
ending at ${\mathbf x}$ were initially located at the left
of the front interface. 
Under very broad conditions, i.e. nonzero molecular diffusivity
and finite variance of the velocity vector potential~\cite{AM,AV,K70}, 
it is possible to show that 
asymptotically the particles undergo a normal diffusion process
with an effective diffusion coefficient 
$D_{\mbox{\scriptsize eff}}$, 
always larger than the molecular value $D_0$. This is the rigorous and
most general version of a statement originally due to Taylor~\cite{Taylor21}.
The issue of single particle diffusion, and the problem of
finding the effective diffusivity,
given a velocity field and a molecular diffusivity, 
has been the subject matter
of a huge amount of work (see e.g.~\cite{MK99} for a recent review). 
In the presence of an asymptotic normal diffusion, we can
substitute the term $\left\langle \theta(0,{\mathbf r}(0) \right\rangle$,
with the gaussian result 
$1-{1 \over 2}\mbox{erfc}(-x/\sqrt{2 D_{\mbox{\scriptsize eff}} t})\simeq 
\exp[-x^2/(4 D_{\mbox{\scriptsize eff}} t)]
/\sqrt{2 \pi D_{\mbox{\scriptsize eff}} t}$, where the latter approximation
holds with exponential accuracy.
We thus obtain $\theta(t,{\mathbf x}) \le 
\exp\left[c_{\mbox{\scriptsize max}}t-
x^2/(4D_{\mbox{\scriptsize eff}}t)\right]
/\sqrt{2 \pi D_{\mbox{\scriptsize eff}} t}$. 
It is therefore clear that at the point ${\mathbf x}$
the field $\theta$ is exponentially small until a time $t$ of the order of
$x/\sqrt{4 D_{\mbox{\scriptsize eff}}c_{\mbox{\scriptsize max}}}$. 
We therefore obtain the upper bound for the front velocity
$V_{f} \le 
\sqrt{4 D_{\mbox{\scriptsize eff}}c_{\mbox{\scriptsize max}}}$,
as anticipated in Eq.~(\ref{eq:upb}).\\
The analytic determination of the effective diffusivity
from the knowledge of the advecting field and the value of the molecular
diffusivity is -- in general -- a daunting task.
Nevertheless there is
an exact result valid for all flows
in the form of an upper bound for the effective diffusivity
$D_{\mbox{\scriptsize eff}} \le D_0(1 + \alpha Pe^2)$, where
$\alpha$ is a numerical constant that depends on the details of the flow
\cite{AM,AV,K70}. 
Plugging this relation into~(\ref{eq:upb}) we can derive a general
upper bound 
\be
V_f/V_0 \le \sqrt{1+\alpha Pe^2}
\label{genupb}
\ee 
where any dependence on the flow details has been
summarized in the numerical constant $\alpha$.
For large $Pe$ we recover the bound $V_f \le {\mbox{const.}}\dot U$ 
already discussed in Ref.~\cite{Const}.
In the limit of small $Pe$, i.e. for small stirring intensity $U$,
this bound is in agreement with the
Clavin-Williams relation $(V_f - V_0)/V_0 = (U/V_0)^2$~\cite{CW}.
On the contrary, in the same limit, an
asymptotic behavior like $(V_f-V_0)/V_0 \sim (U/V_0)^{4/3}$, 
proposed in Ref. \cite{KA},
is ruled out since the rate of convergence to $V_0$ for vanishing
$U$ has to be faster or equal to $U^2$ in order to fulfill
the bound~(\ref{genupb}).\\
As anticipated above, there is a situation where the 
bound~(\ref{eq:upb}) becomes sharp, and that is the
limit of very slow reaction. It is easy to understand the physical
reasons for this effect. If $\tau_r$ is the slowest time-scale
under consideration, advection and molecular diffusion act jointly
to build an effective diffusion process, unaffected by reaction.
Diffusion decreases the value of concentration  to lower levels
before the onset of production, which then takes place  at the maximal
rate (e.g., for FKPP, at $\theta \simeq 0$).
In the limit of very slow reaction, basically one has that 
Eq.~(\ref{eq:rad}), at large scale and long time, behaves as a
reaction-diffusion equation (i.e. with ${\mathbf u}=0$)
where $D_0$ is replaced by $D_{\mbox{\scriptsize eff}}$.
Therefore for FKPP nonlinearity, using Eq.~(\ref{eq:v0}) 
one has: $V_f \simeq 2 \sqrt{\frac{D_{\mbox{\scriptsize eff}}}{\tau_r}}$.
For a detailed derivation of this statement, the reader is referred
to the Appendix~\ref{app:a}.

\section{A discrete-time approach}
\label{sec:3}
Let us now briefly discuss the general idea of our numerical approach to the
study of front dynamics in terms of discrete-time maps.
\\
The physical meaning of Eq.~(\ref{eq:feykacext}) is made clear
by the limit $D_0 = 0$. In that case, introducing
the Lagrangian time derivative
$$ \frac{\mathrm D}{{\mathrm D}t} = \frac{\partial}{\partial t} + 
                                  {\mathbf u} \cdot {\mathbf \nabla} \,\,.$$
Eq.~(\ref{eq:rad}) reduces to
\be
   \frac{\mathrm D}{{\mathrm D}t} \theta = \frac{1}{\tau_r} f(\theta)\,\,.
   \label{eq:radlan}
\ee
Denoting by ${\mathbf F}^t$ the formal evolution operator 
of Eq.~(\ref{eq:langevin}) without noise ($D_0 = 0$), 
i.e. ${\mathbf x}(t) = {\mathbf F}^t {\mathbf x}(0)$,
and by $G^t$ the evolution operator of 
$ \frac{\mathrm d}{{\mathrm d}t} \theta = \frac{1}{\tau_r} f(\theta)$,
i.e. $\theta(t) = G^t \theta(0)$,
one can write the solution of Eq.~(\ref{eq:radlan}) in the form:
\be
   \theta({\mathbf x},t) = G^t \theta({\mathbf F}^{-t}{\mathbf x},0)\,\,.
   \label{eq:exact}
\ee
Equation (\ref{eq:exact}) is nothing but Eq. (\ref{eq:feykacext})
in the absence of molecular diffusivity, i.e. when  only one path
ends in ${\mathbf x}$ at time $t$.\\
In the following, we will concentrate on laminar velocity fields and
we will develop a suitable framework to compute some essential properties
for these systems. In time-periodic velocity fields, 
${\mathbf u}({\mathbf x}, t+\Delta t) = {\mathbf u}({\mathbf x}, t)$ 
where $\Delta t$ is the period and for $D_0 = 0$ the Lagrangian motion 
can be described by a discrete-time dynamical system. 
In other words, the position ${\mathbf x}(t+\Delta t)$ is univoquely
determined by ${\mathbf x}(t)$; in addition, because of the time-periodicity,
the map ${\mathbf x}(t) \to {\mathbf x}(t+\Delta t)$
does not depend on $t$. We remind that a periodic time
dependence is sufficient to induce Lagrangian chaos~\cite{aref}.
\\
With these considerations in mind, we can write a Lagrangian map
for the position:
\be
   {\mathbf x}(t+\Delta t) = {\mathbf F}_{\Delta t}({\mathbf x}(t))\,\,.
   \label{eq:lagrangianmap}
\ee
If ${\mathbf u}$ is incompressible, the map (\ref{eq:lagrangianmap})
is volume preserving i.e. its Jacobian matrix has unit determinant.
In the following we will limit our analysis to the bidimensional case.
In that situation the map (\ref{eq:lagrangianmap}) is symplectic, 
and the dynamics is described by a
discrete-time version of an Hamiltonian system.
Of course it is not simple at all to find explicitly
${\mathbf F}_{\Delta t}({\mathbf x})$ for a generic velocity field. 
On the contrary, it is not difficult to build 
${\mathbf F}_{\Delta t}$ in such a way
that the qualitative features of a given flow are well modeled,
as we will show below.
\\
Another situation in which one obtains exactly a
discrete-time map~(\ref{eq:lagrangianmap})
for the Lagrangian motion, is the case of velocity
field which is always zero apart from $\delta$-impulses at times
$t=0,\pm \Delta t, \pm 2\Delta t, \pm 3\Delta t, \ldots$
\be
 {\mathbf u}({\mathbf x},t) = 
   \sum_{n=-\infty}^\infty {\mathbf v}({\mathbf x}) \delta(t - n\Delta t)\,\,.
 \label{eq:veldimp}
\ee
The effects of a non zero diffusivity are taken into account
by adding a noise term 
\be
   {\mathbf x}(t+\Delta t) = {\mathbf F}_{\Delta t} ({\mathbf x}(t)) + 
                              \sqrt{2D_0 \Delta t\,\,} {\mathbf w}(t)\,\,,
   \label{eq:lagdifmap}
\ee
where ${\mathbf w}(t)$ are standard independent gaussian variables.
\\
If the production term also is zero apart from $\delta$-impulses 
\be
   f(\theta) = \sum_{n=-\infty}^\infty g(\theta) \delta(t - n\Delta t)\,\,,
   \label{eq:reacdimp}
\ee
one can introduce a reaction map
\be
   \theta(t+\Delta t) = G_{\Delta t}(\theta(t))\,\,.
   \label{eq:reactmap}
\ee
Now we are ready to write the dependence of the field $\theta$ at time
$t+\Delta t$ on the field at time $t$ in terms of the advection and
reaction maps, ${\mathbf F}_{\Delta t}$ and
$G_{\Delta t}$
\be
   \theta({\mathbf x}, t+\Delta t) = \left \langle { G_{\Delta t}(
          \theta({\mathbf F}_{\Delta t}^{-1}({\mathbf x} 
          - \sqrt{2D_0 \Delta t\,\,} {\mathbf w}(t)),t))
                                       } \right \rangle_{{\mathbf w}}
   \label{eq:feykacmap}
\ee
Equation~(\ref{eq:feykacmap}) is exactly equivalent to Eq.~(\ref{eq:feykacext})
for maps (for velocity field and reaction given by periodic
$\delta$-impulses).\\ 
The concentration field  
just after the kick, $\theta({\mathbf x}, t + 0)$, 
can be written as 
\be
   \theta({\mathbf x}, t+0) = G_{\Delta t} \left (
        \theta({\mathbf F}_{\Delta t}^{-1}({\mathbf x}), t) \right )\,\,.
   \label{eq:int+0}
\ee
The concentration field $\theta({\mathbf x}, t + \Delta t - 0)$
is obtained from $\theta({\mathbf x}, t + 0)$ solving the bare
diffusion equation $\partial_t \theta = D_0 \nabla^2 \theta$ with
the initial condition given by Eq.~(\ref{eq:int+0}):
\be
   \theta({\mathbf x}, t+\Delta t -0) = \frac{1}{(2\pi)^{d/2}}
       \int e^{-\frac{w^2}{2}} \theta({\mathbf x} - 
       \sqrt{2 D_0 \Delta t}\,{\mathbf w}, t + 0)\, {\mathrm d} {\mathbf w}\,\,,
   \label{eq:fwykacdimpmap}
\ee
which is nothing but Eq.~(\ref{eq:feykacmap}), and $d$ is the
dimension of the space, ${\mathbf x} \in {\rm I}\!{\rm R}^d$.
\\
From an algorithmic point of view the whole process between $t$ and 
$t + \Delta t$ can be thus divided into three steps, a diffusive, an
advective and a reactive one.  The first two steps determine the origin 
of the Lagrangian trajectory ending in ${\mathbf x}$ and accordingly have 
to evolve backwards in time with a given noise realization ${\mathbf w}$.  
In the third step, the reaction at the point ${\mathbf x}$ for the 
advected/diffused passive scalar $\theta$ is computed:
\begin{itemize}
   \item[1)] backward diffusion: 
             ${\mathbf x} \to {\mathbf x} - \sqrt{2D_0 \Delta t}\,
             {\mathbf w}$
   \item[2)] backward advection via the Lagrangian map: ${\mathbf x} -
             \sqrt{2D_0 \Delta t} \, {\mathbf w} \to {\mathbf F}^{-1}
             ({\mathbf x} - \sqrt{2D_0 \Delta t} \, {\mathbf w})$
   \item[3)] forward reaction:
             $\theta(t+\Delta t) = G_{\Delta t}(\theta(t))$.
\end{itemize}
Let us remark that Eq.~(\ref{eq:feykacmap}) is exact if both the
velocity field and the reaction are $\delta$-pulsed processes.
However one can also use the formula (\ref{eq:feykacmap}) as a
practical method for the numerical integration of Eq.(\ref{eq:rad}) if
one assumes small enough $\Delta t$, so that the Lagrangian and reaction maps
are given at the lowest order by $$ {\mathbf F}_{\Delta t}({\mathbf x}) \simeq
{\mathbf x} + {\mathbf u}({\mathbf x}) \Delta t\,\,, \qquad G_{\Delta
t} \simeq \theta + \frac{\Delta t}{\tau_r} f(\theta)\,\,.$$

\subsection{The choice of the reaction map}
\label{sec:3.1}
\begin{figure}[htb]
\epsfig{figure=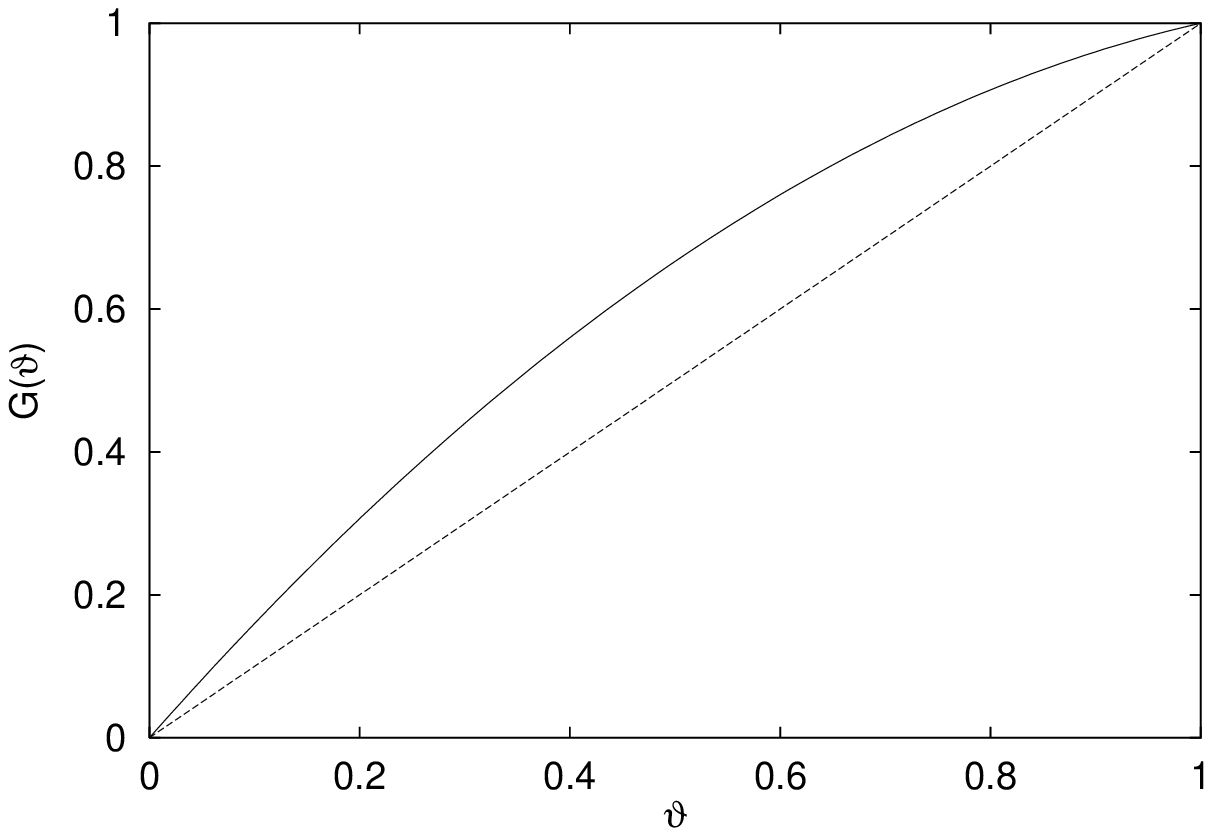,width=0.45\textwidth,angle=0}\hfill
\epsfig{figure=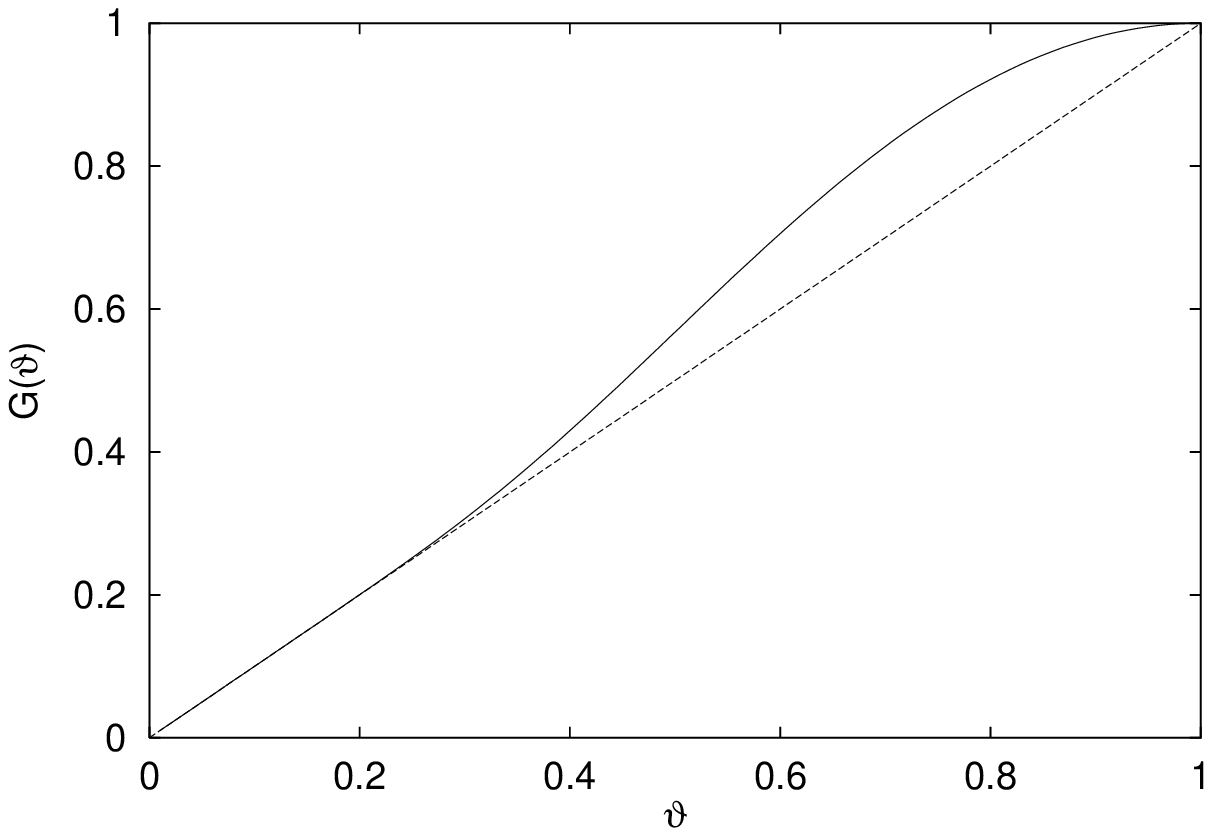,width=0.45\textwidth,angle=0}
\caption{Two examples of reaction maps. On the left FKPP type,
         on the right Arrhenius type.}
\label{fig:reactingmap}
\end{figure}

We now introduce a reaction map $G_{\Delta t}(\theta)$
corresponding to FKPP nonlinearity. The map is
characterized by an unstable fixed point in 
$\theta = 0$, a stable one in $\theta=1$, and a convex shape:
\bea
   G_{\Delta t}(\theta) & = & (1+\alpha \Delta t) \theta + 
	O(\Delta t\,\theta^2) 
 	\qquad {\mathrm {for}} \,\,\, \theta \simeq 0 \nonumber \\
   G_{\Delta t}(\theta) & = & 1 - \beta \Delta t(1-\theta) + O(\Delta t(1-\theta)^2) 
                       \qquad {\mathrm {for}} \,\,\, \theta \simeq 1 \nonumber
\eea
Similarly, for the Arrhenius case we define
\bea
   G_{\Delta t}(\theta) & = & \theta + O(\Delta t\, e^{-\theta_c/\theta}) 
	         \qquad {\mathrm {for}} \,\,\, \theta \simeq 0
\nonumber \\
   G_{\Delta t}(\theta) & = & 1 - \beta \Delta t(1-\theta) + O(\Delta t(1-\theta)^2) 
                 \qquad {\mathrm {for}} \,\,\, \theta \simeq 1\,\,, \nonumber
\eea
(see Figure~\ref{fig:reactingmap}). We expect
from known results~\cite{xin2000}, for the time-continuous
PDE (\ref{eq:rad}) that at a qualitative level,
 the details in the shape of
$G_{\Delta t}(\theta)$ are not very relevant, within a given
 class of nonlinearities (e.g. FKPP).
This expectation is confirmed by numerical simulations.
Naturally, if one is interested in the details of some specific combustion 
processes, one has to work with a precise shape of $G_{\Delta t}$.

\subsection{The choice of the lagrangian map}
\label{sec:3.2}

If we limit our study to the 2D case, the incompressibility of
the velocity field implies symplecticity of the map (\ref{eq:lagrangianmap}).
A rather general class of symplectic maps is the following:
\be
   \left \{ {\begin{array}{l}
                x(t+\Delta t) = x(t) + p_{\Delta t}(y(t))\\
                y(t+\Delta t) = y(t) + q_{\Delta t}(x(t+\Delta t)) \,\,.
             \end{array}
            } \right .
   \label{eq:symplecticmap}
\ee
It is easy to verify that (\ref{eq:symplecticmap}) is symplectic
for any choice of $p_{\Delta t}(\cdot)$ and $q_{\Delta t}(\cdot)$. 
If $p_{\Delta t}(y) = k\,\Delta t\, \sin y$ and 
$q_{\Delta t}(x) = k\,\Delta t\, \sin x$ one has 
the so called Harper model (often studied in the quantum chaos
context) corresponding to a chaotic transport in spatially periodic
cellular structures.
The case $q_{\Delta t}(x)=0$ gives a non-chaotic 
shear flow in the $x$ direction.
The celebrated standard map~\cite{LichLieb}, which is a paradigmatic 
model for chaotic behaviour in low dimensional Hamiltonian system,
is obtained with $p_{\Delta t}(y) = k\,\Delta t\, \sin y$ and $q_{\Delta t}(x)=x$.
\\
In the following we will study two limit cases which give 
nonchaotic transport, i.e. the role of the molecular diffusivity
is very important:
\begin{itemize}
   \item[a)] open flow field (shear flow) where all the streamlines
             are open
   \item[b)] convective rolls where all the streamlines are closed.
\end{itemize}
For the shear flow, we set
$q_{\Delta t}(x)=0$ in  (\ref{eq:symplecticmap}), as mentioned above. 
For the cellular flows we have built the
Lagrangian map in the following way: consider a 2d incompressible steady
velocity field ${\mathbf u}(x,y) = (-\partial_y \psi(x,y), \partial_x
\psi(x,y))$ generated by the streamfunction
\be
   \psi(x,y) = UL\sin(\frac{2\pi x}{L})\sin(\frac{2\pi y}{L}) \,\,,
   \label{eq:streamfunction}
\ee
with $L$-periodic conditions in $y$ and infinite extent along the $x$-axis.
The Lagrangian map ${\mathbf F}_{\Delta t}({\mathbf x})$ is given
by the exact integration of the equation 
$\frac{\mathrm d}{{\mathrm d}t} {\mathbf x} = {\mathbf u}({\mathbf x})$ 
on an interval $\Delta t$.
The shape of ${\mathbf F}_{\Delta t}({\mathbf x})$, i.e.
the expression of ${\mathbf x}(t+\Delta t)$ as a function of ${\mathbf x}(t)$
is found explicitly in terms of elliptic functions.
\\
In addition to the cases a) and b) we will study the relevance of
a ``transversal'' perturbation to the shear flow. 

\section{Numerical results}
\label{sec:4}
Since we are interested in the front propagation in one direction, say
 the $x$-direction, we applied in our simulations
periodic boundary conditions in $y$-direction:
\begin{equation}
\label{bc-y}
\theta(x,y,t)=\theta(x,y+L_y,t)
\end{equation}
whereas in $x$-direction we have
\begin{eqnarray}
\label{bc-x}
\lim_{x\rightarrow \infty}\theta(x,y,t)&=&0\;,\\
\lim_{x\rightarrow -\infty}\theta(x,y,t)&=&1 \;.
\end{eqnarray}
In this way the front propagates from left to right.  
\\ 
The
instantaneous front speed $V(t)$ is defined by
\begin{equation}
V(t) = {1\over {\Delta t L_y}}\int\int dx\,dy
[\theta(x,y,t+\Delta t)-\theta(x,y,t)]\;.
\end{equation}
Since $V(t)$ in general shows some oscillations in time, one
can define the mean front speed $V_f$ by the time average over a
sufficiently long time, after the transient.
\\
The numerical implementation of the formula (\ref{eq:feykacmap}) 
is described in detail in Appendix C. 
\\
We have first checked the numerical code to confirm the known results
for the front speed $V_f$ in the case of  FKPP nonlinearity:
\begin{itemize}
   \item[i)] only molecular diffusion is present, i.e. ${\mathbf F(x)=x}$.
	     In this case, equivalent to ${\mathbf u}={\mathbf 0}$,
             one has the discrete time version of the FKPP 
	     formula~(\ref{eq:v0})
             \be
                 V_f = \frac{2}{\Delta t} \sqrt{D_0\Delta t \, 
                        \ln \left ( G'_{\Delta t}(0) \right )}\,\,.
                 \label{eq:vfdiscrkpp}
             \ee
             See Appendix B for its derivation.
   \item[ii)] the reaction is very slow, i.e., 
              $\tau_r = \Delta t / \ln (G'_{\Delta t}(0))>> t_a$ 
              (where $t_a$ is the advection time). In this case homogenization 
              techniques can be applied (see e.g. \cite{Const}, and
	      Appendix~\ref{app:a}) and one finds
              \be
                  V_f \simeq \frac{2}{\Delta t} \sqrt{D_{\mbox{\scriptsize eff}}\,\Delta t\,
                        \ln \left ( G'_{\Delta t}(0) \right )}
                        = 2\sqrt{D_{\mbox{\scriptsize eff}}\over\tau_r}\;.
	          \label{eq:homog}
              \end{equation}
\end{itemize}
The effective Diffusion coefficient $D_{\mbox{\scriptsize eff}}$ 
in the $x$-direction is defined by
\begin{equation}
D_{\mbox{\scriptsize eff}} = \lim_{t \to \infty}
\frac{\langle(x(t)-x(0))^2\rangle}{2t}
\end{equation}
and in general must be computed numerically iterating the map 
(\ref{eq:lagdifmap}).
\\
In order to show the validity of the homogenization limit 
(\ref{eq:homog}) we consider a system where the lagrangian
motion is given by the standard map
\be
   \left \{ {\begin{array}{lll}
                x(t+1) & = & x(t) + K\sin(y(t))\\
                y(t+1) & = & y(t) + x(t+1) \;\;\;\;\mbox{mod} \,\,\,2\pi\,\,,
             \end{array}
            } \right .
   \label{eq:standard}
\ee
and the reaction map $G(\theta)$ is in the FKPP class
\be
   G(\theta) = \theta + c \theta (1 - \theta) \qquad \mbox{if $c \leq 1$}\,\,.
   \label{eq:FKPP1}
\ee
In the case of $c>1$, in order to avoid unbounded reaction map, we use
\be 
   G(\theta) = \left \{ {\begin{array}{ll}
                \theta + c \theta (1 - \theta) & \qquad 
	 \mbox{for} \qquad \theta \leq \theta^*\\
                1 - \alpha (1 - \theta) & \qquad 
         \mbox{for} \qquad \theta > \theta^*\,\,,
             \end{array}
            } \right .
   \label{eq:FKPP2}
\ee
where $\theta^* = (c + 2 - \sqrt{c^2+4})/2c$
and $\alpha = (c + 2 - \sqrt{c^2+4})/(c - 2 + \sqrt{4+c^2})$.

\begin{center}
\begin{figure}[htb]
\epsfig{figure=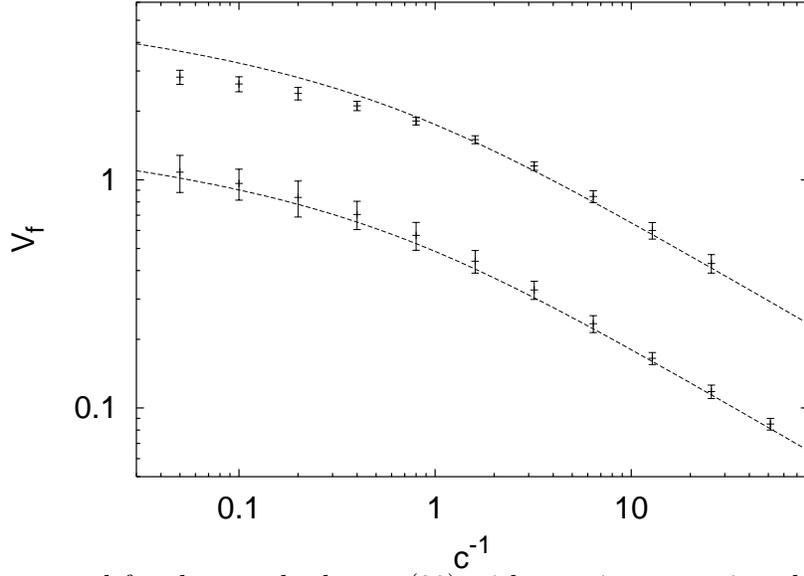,width=0.7\textwidth,angle=0}\hfill
\caption{Front speed for the standard map (\ref{eq:standard})
with reaction map given by Eq.~(\ref{eq:FKPP1}) (\ref{eq:FKPP2}),
as function of $c$, $D_0 = 0.04$.
The upper curve is for $K=3.0$, the lower for $K=1.0$. 
The dotted lines are the homogenization curves
$2\sqrt{D_{\mbox{\scriptsize eff}} \ln(1.0 + c)}$.
The diffusion coefficient $D_{\mbox{\scriptsize eff}}$ 
depends on $K$ and has been computed numerically.} 
\label{fig:homostand}
\end{figure}
\end{center}

By varying $c$ one can change the ratio between
the advection time $t_a \sim O(1)$ and the reaction time
$\tau_r = \frac{1}{\ln G'(0)} = \frac{1}{\ln(1+c)}$, i.e. the 
Damk\"ohler number.
Figure~\ref{fig:homostand} shows $V_f$ vs $1/c$, for $K = 1$ 
and $K = 3$. At large $\tau_r$, i.e. large $c^{-1}$ 
the homogenization limit is recovered.
\\
We now move to the description of the numerical results in three
case studies.
\subsection{Shear Flow}
\label{sec:4a}

The shear flow is the simplest case to study and it will be
presented here shortly. 
For our simulations we use the system (\ref{eq:symplecticmap}) with
$q=0$:
\be
   \left \{ {\begin{array}{l}
                x(t+1) = x(t) + U\sin (2 \pi y(t)/L_y)\\
                y(t+1) = y(t) \,\,.
             \end{array}
            } \right .
   \label{eq:shearmap}
\ee
and a reaction map $G(\theta)$
of the shape (\ref{eq:FKPP1}) (\ref{eq:FKPP2}), where
for the sake of simplicity we use $\Delta t=1$.
While keeping fixed the diffusivity we
investigate the different regimes in the $(U,\tau_r)$ space.  
Among all the possible combinations of diffusion, advection
and reaction time scales, we assume in this paper always that diffusion is the
slowest one.\\
The front velocity in the homogenization regime, i.e. very slow reaction, 
is $V_f \simeq 2\sqrt{D_{\mbox{\scriptsize eff}}\over \tau_r}$,
where $D_{\mbox{\scriptsize eff}}$ can be easily computed for the
shear flow~\cite{Zeldovich}: 
$D_{\mbox{\scriptsize eff}} - D_0 \sim U^2/D_0$. Therefore for
slow reaction one has a linear behaviour $V_f \sim U$.
The bounds discussed in the introduction suggest
that in general $V_f=aU+b$ where $a$ and $b$ may 
depend on $\tau_r$ and $D_0$, as confirmed by
recent numerical results \cite{vladimirova}.  

\begin{center}
\begin{figure}[htb]
\epsfig{figure=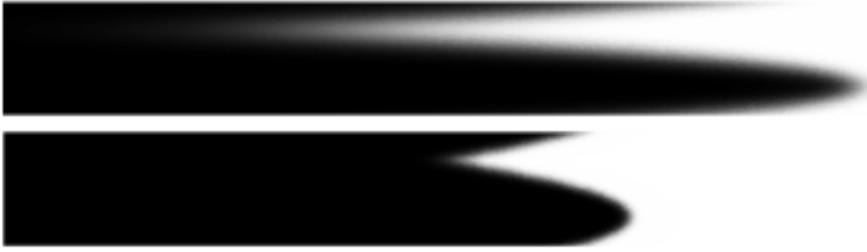,width=0.95\textwidth,angle=0}\hfill
\caption{Snapshots of the field $\theta(x,y)$
for the shear map (\ref{eq:shearmap}) and the reaction map 
(\ref{eq:FKPP1},\ref{eq:FKPP2}).  $U = 0.5$, $D_0 = 0.01$,
$c = 0.2$ and $c = 2.0$ for the upper and lower image
respectively. The system size is $L_y = 2\pi$ and $L_x = 20\pi$.}
\label{fig:snapshear}
\end{figure}
\end{center}

\begin{center}
\begin{figure}[htb]
\epsfig{figure=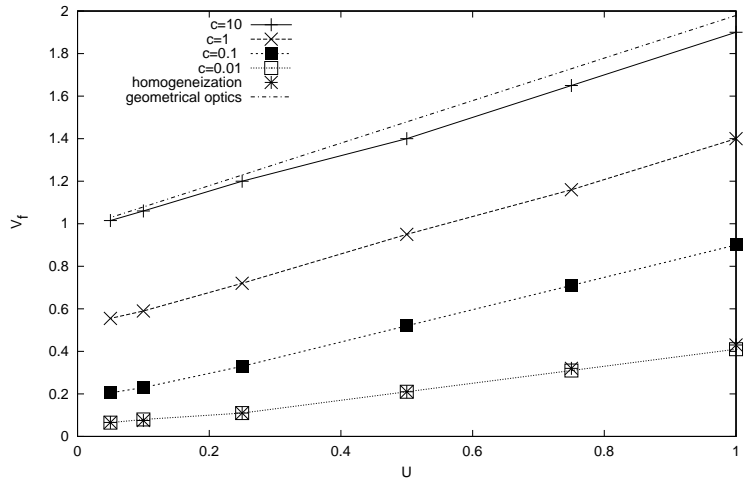,width=0.6\textwidth,angle=0}\hfill
\caption{Front speed for the shear map (\ref{eq:shearmap}) 
and the reaction map (\ref{eq:FKPP1},
\ref{eq:FKPP2}) as function of $U$ for various reaction times
$\tau_r = {1 \over \ln(1+c)}$, $D_0 = 0.04$.} 
\label{fig:shear1}
\end{figure}
\end{center}

In Fig.~\ref{fig:snapshear} we show two snapshots of the concentration
field for slow and fast reaction.
In Fig.~\ref{fig:shear1} the front velocity is displayed in dependence
on the advection velocity $U$ for different reaction times.
homogeneization holds for slow reaction rates; decreasing $\tau_r$ the 
front speed increases until for high reaction rates the geometrical optics
regime is reached. 

Our results obtained with the discrete time map approach are in perfect
agreement with the direct numerical simulations presented in
Ref.~\cite{vladimirova}, where the dynamical equations are solved 
in the Eulerian framework.

\subsection{Cellular Flow}
\label{sec:4b}
\begin{center}
\begin{figure}[htb]
\epsfig{figure=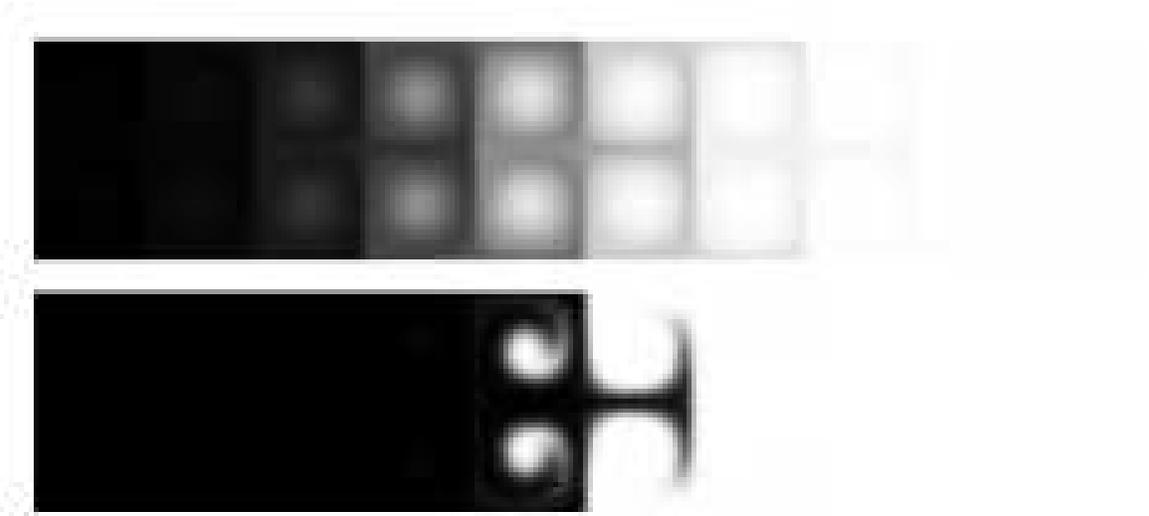,width=0.95\textwidth,angle=0}\hfill
\caption{Snapshots of the field $\theta(x,y)$ for cellular 
flow given by the streamfunction~(\ref{eq:streamfunction})
and the reaction map (\ref{eq:FKPP1}, \ref{eq:FKPP2}),
$U = 2.0$, $D_0 = 0.01$, $\tau_r = 5.0$ and $\tau_r = 0.5$ 
for the upper and lower image respectively. 
The system size is $L_y = 2\pi$ and $L_x = 10\pi$.}
\label{fig:snapgol}
\end{figure}
\end{center}

The numerical simulations have been performed using the velocity field
${\mathbf u}(x,y) = (-\partial_y \psi(x,y), \partial_x \psi(x,y))$
generated by the streamfunction defined in Eq.~(\ref{eq:streamfunction}) and
$G(\theta)$ given by (\ref{eq:FKPP1}) and (\ref{eq:FKPP2}) with 
$c=\Delta t/\tau_r$ and $\Delta t \ll \tau_r$. In 
Fig.~\ref{fig:snapgol} we show a snapshot of the concentration field
for two values of $\tau_r$.

The key to the understanding of the different regimes of front 
propagation stands in the description of front dynamics in terms 
of effective macroscopic equations, which we introduce hereafter,
following Ref.~\cite{Pomeau}.
The dynamics of $\theta$ is characterized by the length-scale
of the cell-size, $L$. We can therefore perform a space
discretization which reduces each cell, $C_i$, to a point, $i$, 
mapping the domain -- a two-dimensional 
infinite strip -- to a one-dimensional lattice,
and the field $\theta$ to a 
function defined on the lattice $\Theta_i = L^{-2} \int_{C_i} \theta \,dx\,dy$.
Integrating Eq.~(\ref{eq:rad}) over the cell $C_i$ we obtain 
$\dot{\Theta}_i = J_{i+1}-J_{i} +\chi_i$ where 
$J_{i}=L^{-2}\int_{\mbox{\scriptsize left}} D_0 \partial_x\theta \,dy$ is the flux of matter
through the left boundary of the $i$-th cell, and 
$\chi_i=L^{-2}\int_{C_i} \tau_r^{-1} f(\theta) \,dx\,dy$ is the rate of change of 
$\Theta_i$ due to reaction taking place within the cell.
We will show that it is possible to model the dynamics
with a space-discretized macroscopic reaction-diffusion equation:
\begin{equation}
\dot{\Theta}_i= D_{\mbox{\scriptsize eff}}(\frac{1}{2} \Theta_{i+1} -\Theta_i + 
{1 \over 2}\Theta_{i-1}) + 
\tau_{\mbox{\scriptsize eff}}^{-1} F(\Theta_i)\, .  
\label{eq:3}
\end{equation}
The effect of velocity is to {\em renormalize} the values
of diffusivity, $D_0 \to D_{\mbox{\scriptsize eff}}(D_0,U,L)$, and reaction time-scale, 
$\tau_r \to \tau_{\mbox{\scriptsize eff}}(\tau_r,U,L)$ and therefore the advective term does not
appear any longer in the effective dynamics given by Eq.~(\ref{eq:3}).
The assumption that $\kappa$ and $\tau$ are
independently renormalized by advection is consistent in the
regime $v/U = (Da/Pe)^{1/2} \ll 1$

\begin{figure}
\centerline{\includegraphics[scale=0.3,draft=false]{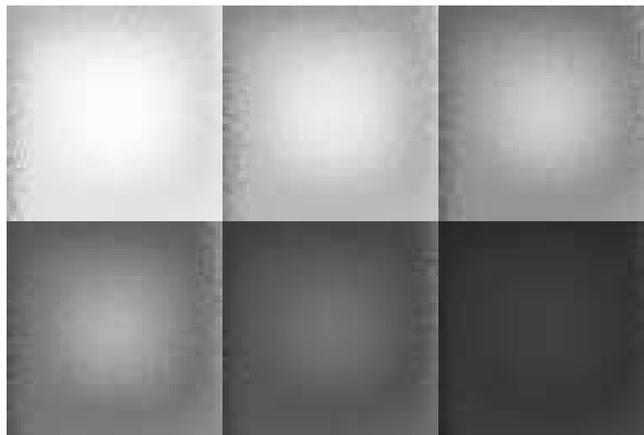}}
\caption{Cellular flow: six snapshots of the field $\theta$ within the
same cell, at six successive times with a delay $\tau/6$ 
(from left to right, top to bottom), as a result of
the numerical integration of equation~(\protect\ref{eq:rad}).
Here $Da \simeq 0.4,Pe\simeq 315$. Black stands for
$\theta=1$, white for $\theta=0$.}
\label{fig:snap1}
\end{figure}

The renormalized diffusivity $D_{\mbox{\scriptsize eff}}$ accounts for the process of diffusion
from cell to cell resulting from
the nontrivial interaction
of advection and molecular diffusion. The renormalized reaction-time
$\tau_{\mbox{\scriptsize eff}}$ amounts to the time that it takes to a single
cell to be completely burned, and depends on the interaction of advection and
production.  
In that context, the limiting speed of the front 
in the moving medium will be 
$V_f \simeq 2\sqrt{D_{\mbox{\scriptsize eff}}/\tau_{\mbox{\scriptsize eff}}}$.
The goal is now to derive 
the expressions for the renormalized parameters from physical considerations.\\
{\em Renormalization of diffusivity.\/} 
To obtain the value of $D_{\mbox{\scriptsize eff}}$ it is sufficient to neglect 
the reaction term in equation~(\ref{eq:rad}), i.e. consider a passive scalar
in a cellular flow. The solution is known, \cite{Ros,Shr,Pom}, 
\begin{equation}
{D_{\mbox{\scriptsize eff}}\over D_0} \sim
Pe^{1/2} \qquad Pe \gg 1 \;.
\label{eq:4}
\end{equation}
For large $Pe$ ($D_0$ small)
the cell-to-cell diffusion mechanism can be qualitatively understood
in the following way: the probability for a particle of the scalar to jump
across the boundary of the cell, in a circulation time $L/U$, by virtue
of molecular diffusion, can be estimated as the ratio of the diffusive motion 
across the streamlines, $O(\sqrt{D_0 L/U})$, 
to advective motion along streamlines, 
$O(L)$, leading to $p\sim(D_0/(UL))^{1/2}$, hence the effective diffusivity
$D_{\mbox{\scriptsize eff}}\sim p\,UL \sim D_0 Pe^{1/2}$.\\
{\em Renormalization of reaction time.\/} 
At small $Da$, where
reaction is significantly slower than advection, the 
cell is first invaded by a mixture of reactants and products
(with a low content of products, $\Theta_i \ll 1$) on the
fast advective time-scale,
and complete reaction ($\Theta_i=1$) 
is then achieved on the slow time-scale 
$\tau_{\mbox{\scriptsize eff}}\simeq\tau_r$ 
(Figure~\ref{fig:snap1}). 
The area where the reaction takes place extends over several cells,
i.e. the front is ``distributed''.

\begin{figure}
\centerline{\includegraphics[scale=0.3,draft=false]{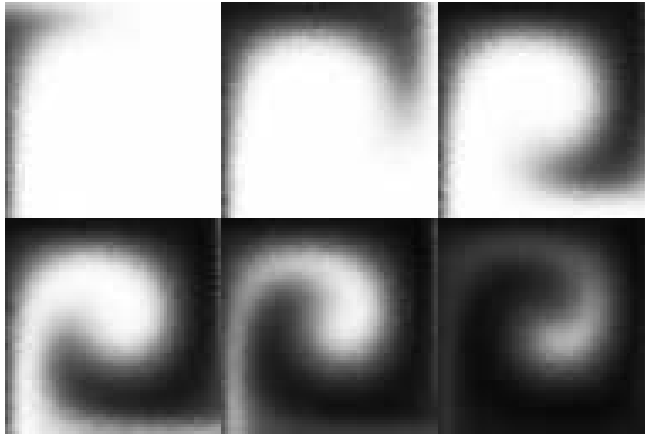}}
\caption{Cellular flow: six snapshots of the field $\theta$ within 
the same cell, at six successive times with a delay $(L/U)/6$ 
(left to right, top to bottom). Here $Da=4,Pe=315$. A spiral wave 
invades the interior of the cell, with a speed comparable to $U$.}
\label{fig:snap2}
\end{figure}

\begin{figure} 
\centerline{\includegraphics[scale=0.45,draft=false,angle=270]{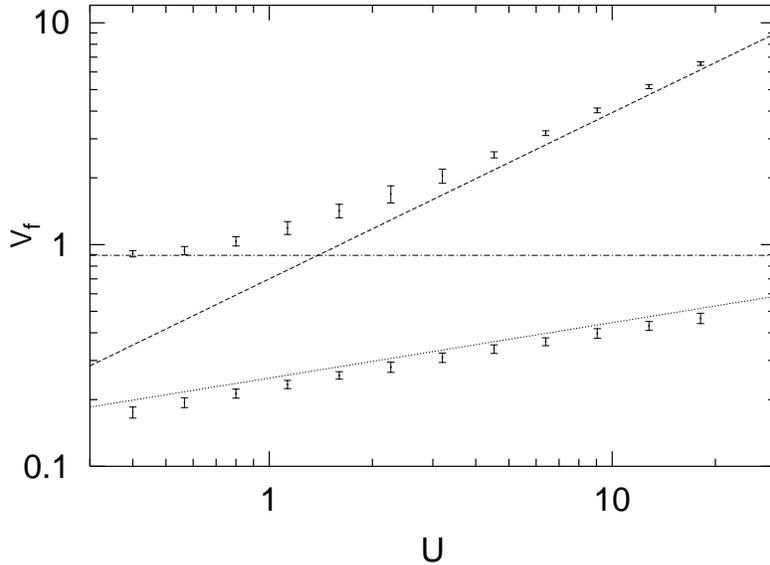}}
\caption{Cellular flow: the front speed $V_f$ as a function of $U$, 
the typical flow velocity, with $D_0 = 0.04$. The lower curve shows
data at $\tau_r = 20.0$ (fast advection).
The upper curve shows data at $\tau_r = 0.2$ (slow advection). For comparison,
the scalings $U^{1/4}$ and $U^{3/4}$ are shown as dotted and dashed
lines respectively. The horizontal line indicate $V_0$ (the front 
speed without advection, i.e. $U=0$) for $\tau_r = 0.2$.}
\label{fig:ufgoll}
\end{figure}

At large $Da$, the ratio of time-scales reverses, and in a (now short) 
time $\tau_r$ 
two well-separated phases emerge
inside the cell. The interface has a depth 
$\lambda \sim \sqrt{D_0\tau_r} =L Pe^{-1/2} Da^{-1/2}$, i.e. it is thin 
compared to the cell size. Here the process is characterized 
by an inward spiral motion of the outer, stable phase 
(see Figure~\ref{fig:snap2}), at a speed proportional 
to $U$ as it usually happens for a front in a shear flow at large $Da$. 
Indeed it is easy to show that, inside a cell, the problem can be
mapped to a front propagation in a shear-flow in
``action-angle'' variables \cite{Rhi}.
Therefore the $\theta=1$ phase fills the whole cell 
on the advective time-scale, giving $\tau_{\mbox{\scriptsize eff}}\simeq L/U$.\\
In summary, we have the following behavior for the renormalized
reaction time
\begin{equation}
{\tau_{\mbox{\scriptsize eff}}\over \tau_r} \sim \left\{
{\begin{array}{lll}
1 & & Da \ll 1 \;\,\, \\
Da & & Da \gg 1 \;.
\end{array}}  \right.
\label{eq:5}
\end{equation}
Now, we have all the information to derive the effective speed
of front propagation for a cellular flow. 
Recalling that $V_f \sim \sqrt{D_{\mbox{\scriptsize eff}}/\tau_{\mbox{\scriptsize eff}}}$, 
we have for the front velocity the final result
\begin{equation}
{V_f\over V_0} \sim \left\{
{\begin{array}{lll}
Pe^{1/4} & & Da \ll 1, Pe \gg 1 \;\,\, \\
Pe^{1/4}Da^{-1/2} & & Da \gg 1, Pe \gg 1 
\end{array}}  \right .
\label{eq:6}
\end{equation}
where we restricted ourselves to the most interesting case $Pe \gg 1$.
At small $Da$ the front propagates with an effective velocity which
scales as the upper bound derived above, that is as $Pe^{1/4}$.
At large $Da$ front speed is less enhanced than at small $Da$: according
to Eq.~(\ref{eq:6}), we have 
$V_f/\sqrt{4D_{\mbox{\scriptsize eff}}/\tau_r} \sim Da^{-1/2}$ 
for $Da \gg 1$. 
In terms of the typical velocity of the cellular flow, we have
$V_f \propto U^{1/4}$ for ``fast'' advection 
($U \gg L/\tau_r$, or equivalently $Da \ll 1$) whereas
$V_f \propto U^{3/4}$ for ``slow'' advection 
($U \ll L/\tau_r$, or $Da \gg 1$). 
The numerical results are shown in Figure~\ref{fig:ufgoll}.
The case of ``fast'' advection corresponds to the one
with slow reaction, for which the homogenization limit holds.

\begin{figure} 
\centerline{\includegraphics[scale=0.4,draft=false,angle=270]{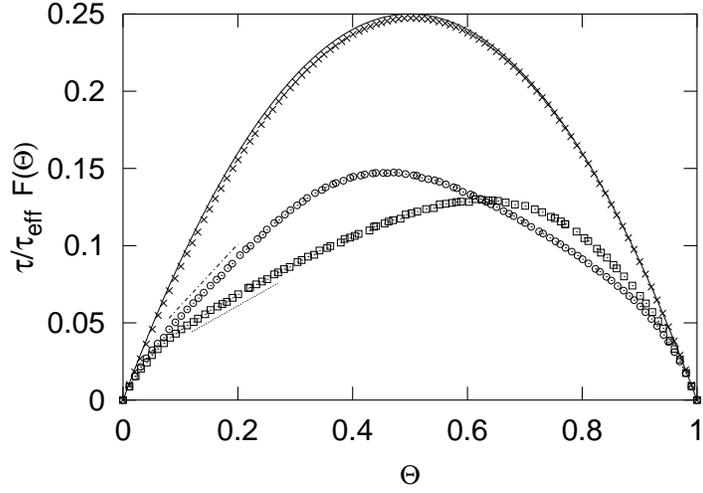}}
\caption{Cellular flow: the renormalized reaction term 
$\tau_r / \tau_{\mbox{\scriptsize eff}}\, F(\Theta)$ vs $\Theta$ 
for three different parameters:
$Da\simeq 4$ ($\Box$), $Da\simeq 2$ ($\circ$) 
and $Da \simeq 0.4 $ ($\times$).
The continuous line is $f(\theta)$. 
The dotted and dash-dotted lines have the slopes ($0.2$ and $0.4$) 
proportional to $Da^{-1}$ 
in the region of slow advection.}
\label{fig:ftheta}
\end{figure}

In the geometrical optics limit $Da \gg 1, Pe \gg 1$, 
the effective speed of the front is proportional 
to the area of the interface
that separates the two phases. In two dimensions, the interface is 
characterized by its length, $\ell$, and its depth, $\lambda$.
We have the relationship $V_f \sim  \lambda \ell / (L \tau_r)$
which entails the result that the ratio of the length
of the interface in a moving medium, $\ell$, 
to the length in a medium at rest, $L$, is $\ell/L \sim V_f/V_0$ and is
therefore larger than unity. 
The structure responsible for this elongation of the 
front edge is the spiral wave shown in Figure~\ref{fig:snap2}.\\
Finally, it is interesting to look at the shape of the effective reaction
term $\tau_{\mbox{\scriptsize eff}}^{-1} F(\Theta)$ appearing in the
renormalized equation~(\ref{eq:3}). As shown in Figure~\ref{fig:ftheta},
for small $Da$ the effective production term is indistinguishable from
the ``bare'' one. Increasing $Da$, the reaction rate
tends to reduce progressively,  inducing the slow-down of the front speed. 
The effective potential shows a small region where
the production term is essentially the microscopic one, followed by an
intermediate regime characterized
by a linear dependence on the cell-averaged concentration, 
with a slope directly proportional to $Da^{-1}$. 
That is in agreement with a typical effective reaction time 
$\tau_{\mbox{\scriptsize eff}}\sim \tau_r Da$
(cf. Eq.~(\ref{eq:5})). 

We conclude the discussion on cellular flows by noting that the scaling 
behaviors $V_f$ vs $U$ are in agreement with the rigorous bounds
$V_f \geq C_1 U^{1/5} + C_2$ and $V_f \geq C_3 U^{1/2} + C_4$ 
for slow and fast reaction~\cite{KR00}.\\

\subsection{Percolating Flow}
\label{sec:4c}
In the previous subsections we discussed pure cellular and 
pure shear flows. Now we investigate the transition
between these two limiting cases.
To this aim, we will use for the lagrangian motion
the generalized Harper map:
\be
   \left \{ {\begin{array}{l}
                x(t+1) = x(t) + U\sin y(t)\\
                y(t+1) = y(t) + U_T\sin x(t+1)\,\,\,\,\mbox{mod}
					      \,\,\,\, 2\pi\,.
             \end{array}
            } \right .
   \label{eq:harper}
\ee
The case $U_T=0$ corresponds to shear flow, whereas
$U_T = U$ gives a chaotic cellular flow.

\begin{center}
\begin{figure}[htb]
\epsfig{figure=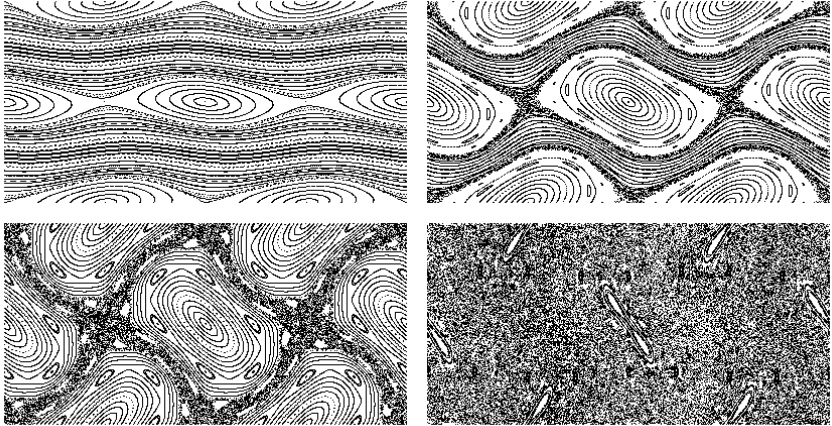,
	width=0.7\textwidth,angle=0}\hfill
\caption{Lagrangian dynamics of test particles evolving according to
the Harper's map (\protect{\ref{eq:harper}}) for different values of $U_T$,
with $U=1.5$. From top-left moving clockwise:
$U_T = 0.2$, $U_T = 0.8$, $U_T = 1.5$ and $U_T = 3.0$.}
\label{fig:harpmap}
\end{figure}
\end{center}

\begin{center}
\begin{figure}[htb]
\epsfig{figure=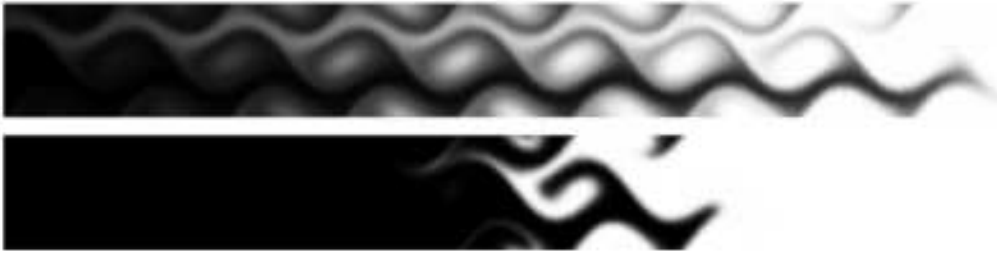,
	width=0.95\textwidth,angle=0}\hfill
\caption{Snapshots of the field $\theta(x,y)$
for percolating flow given by Eq.~(\ref{eq:harper}).
The reaction map is given by (\ref{eq:FKPP1},\ref{eq:FKPP2}). 
$U = 1.5$, $U_T = 0.8$, $D_0 = 0.01$, $c = 0.2$ 
and $c = 2.0$ for the upper and lower image
respectively. The system size is $L_y = 2\pi$ and $L_x = 20\pi$.}
\label{fig:snapharp}
\end{figure}
\end{center}

\begin{center}
\begin{figure}[htb]
\epsfig{figure=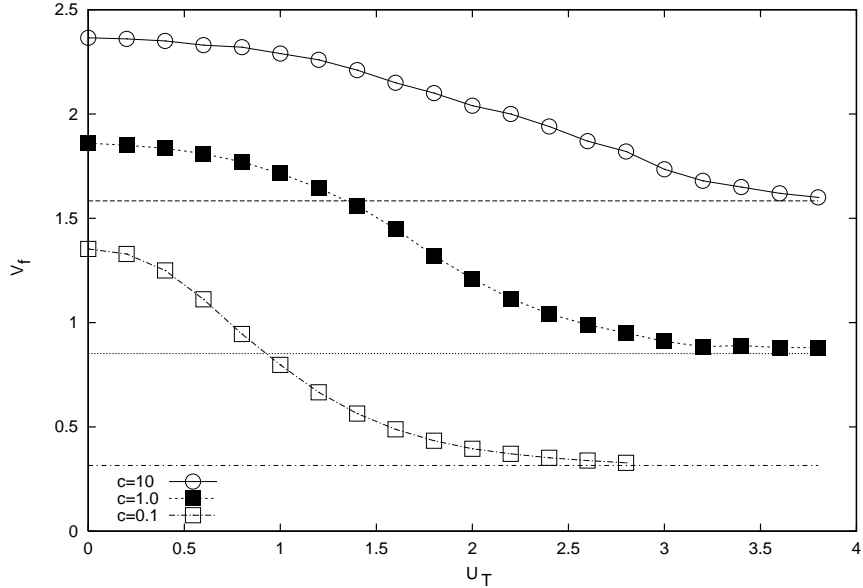,width=0.7\textwidth,angle=0}\hfill
\caption{$V_f$ vs $U_T$, the Lagrangian map is given by 
(\ref{eq:harper}), the reaction map $G(\theta)$ is given by 
(\ref{eq:FKPP1},\ref{eq:FKPP2}).
The three curves correspond to $c=0.1,1.0,10.0$ (from bottom to top),
the ``horizontal'' velocity has been fixed to $U=1.5$ for all
curves. The asymptotic value of
$2\sqrt{D_{\mbox{\scriptsize eff}}\ln(1+c)}$ (for the horizontal
direction) is shown by the three horizontal lines, the corresponding
value of $D_{\mbox{\scriptsize eff}}$ has been calculated numerically
for large $U_T$.}
\label{fig:percolation} 
\end{figure} 
\end{center}

In order to give an idea of the Lagrangian behaviour of the 
flow generated by~(\ref{eq:harper}) we show in Fig.~\ref{fig:harpmap}
some trajectories at different
value of $U_T$. For $U_T \neq 0$ the map~(\ref{eq:harper})
exhibits chaotic behaviour in some regions. At small value
of $U_T$ one has basically a ballistic transport in the $x$ direction
apart from small recirculation regions.
For $U_T \sim U$ a typical ``cat's eye'' pattern appears with
percolating channels among the re-circulation regions. 
A chaotic cellular flow, rather
similar to the case of convective cells discussed in section IV B 
(apart from a rotation of $\pi/4$) is obtained
for $U_T = U$, see Fig.~\ref{fig:harpmap}. The behaviour for
$U_T \gg U$ can be understood by a simple statistical
argument valid for large $K$~\cite{LichLieb}:
at very large values of $U_T$, $y(t)$ changes very rapidly,
therefore the oscillatory term $\sin(y(t))$ 
can be considered as a zero mean random process
and the variable $x(t)$ is well approximated
by a diffusive process with $D_{\mbox{\scriptsize eff}}\simeq U^2/4$.
\\
In Fig.~\ref{fig:snapharp} we show snapshots of the concentration field
for different reaction times.
The natural question is how the transition from pure shear to
percolation and from cellular flow to percolation, respectively,
changes the front speed.
On an intuitive basis we expect the front
propagation  to be slower in the cellular case than in the shear case. 
In Fig.~\ref{fig:percolation} we plot the front speed in
dependence of the sidewind $U_T$ for different reaction time scales. 
The diffusion coefficient of the system (\ref{eq:harper})
has been calculated in separate runs for
the comparison with the homogeneization expectations.

Let us discuss the figure going from left to right, augmenting the
sidewind. For zero sidewind, we recover the pure shear result, at the
value $U_T=U$, we recover the pure cellular flow result and if we go
even beyond, for a very large sidewind, the reaction becomes
relatively small and thus we enter the homogeneization regime. This
transition pure shear-pure cells is smooth as confirmed by the figure. 
The three horizontal lines show the asymptotic values of
$2\sqrt{D_{\mbox{\scriptsize eff}}/\tau_r}$ 
(the value of $D_{\mbox{\scriptsize eff}}$ does not change
significantly as a function of $U_T$ if $U_T$ is large enough).

\subsection{Final remarks}
\label{sec:4d}
\begin{figure} 
\centerline{\includegraphics[scale=0.4,draft=false,angle=270]{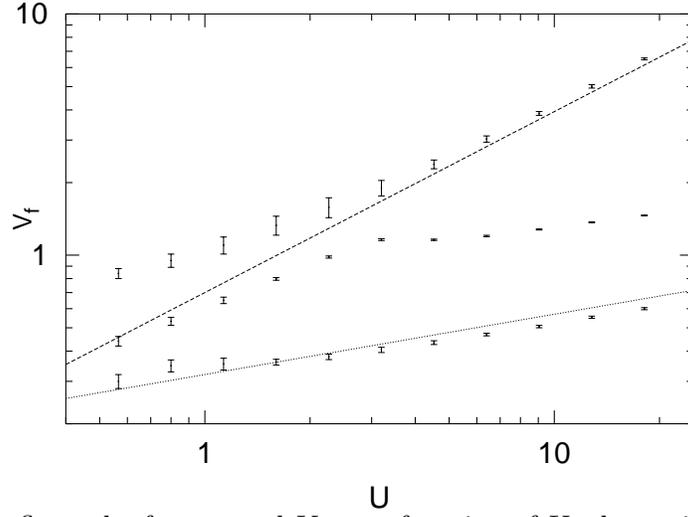}}
\caption{Cellular flow: the front speed $V_f$ as a function of $U$, 
the typical flow velocity in the case of Arrhenius production type: 
$f(\theta) = (1-\theta)\exp(-\theta_c/\theta)$. The lower curve shows 
data at $\tau_r = 2.0$ and $\theta_c = 0.5$ (fast advection). 
The intermediate curve at $\tau_r=2.0$ and $\theta_c = 0.2$
shows the crossover from fast (right side) to slow advection 
(left side).
The upper curve shows data at $\tau_r = 0.2$ and $\theta_c = 0.2$
(slow advection). For comparison, the scalings $U^{1/4}$ and 
$U^{3/4}$ are shown as dotted and dashed lines respectively.}
\label{fig:arrenius}
\end{figure}

In the previous subsections we have shown the results for the behaviour
of $V_f$ as function of $U$ and $\tau_r$ for different laminar
flows discussing in detail the FKPP nonlinearity.
It is natural to ask about the effects of the shape
of $f(\theta)$ on the front speed $V_f$. In particular
it is interesting to know whether the choice of an Arrhenius nonlinearity
changes significantly the scenario presented in the previous sections.
\\
It is known that for ignition
nonlinearity, i.e. $f(\theta)=0$ for $\theta < \theta_c$,
and expectedly also in the Arrhenius case,
the flow can suppress front propagation. This effect, called flame quenching,
is absent for FKKP production terms.
This observation may lead to the assumption that 
 the front evolution could depend 
on the shape of $f(\theta)$ in a dramatic way.
However, flame quenching takes place only if  initial conditions
of the field $\theta$ are localized, i.e. $\theta$ is different
from zero only in a bounded region \cite{quenching}. 
For the initial and boundary
conditions that we use here (Eqs. (\ref{bc-x})), 
the front propagates always~\cite{Const,KR00}, also in the Arrhenius case.
For that reason, we do not expect major differences in the scaling
properties of propagation speeds. Indeed,
in the particular geometry we use, i.e. an open flow 
with an infinite reservoir of burned material,
the values of $V_f$ in the case of Arrhenius nonlinearity
are very similar to the ones obtained using the FKPP
nonlinearity.
In Fig.~\ref{fig:arrenius} we show $V_f$ as a function of $U$ 
for the cellular 
flow introduced in  Section IV B in the case of Arrhenius nonlinearity. 
The scaling laws
$V_f \sim U^{1/4}$ and $V_f \sim U^{3/4}$ for slow
and fast reaction hold also in this case. 
Also for shear and percolating flow we do not
observe qualitative changes when varying the shape
of $f(\theta)$.

Although the qualitative behaviour of $V_f$ as a function 
of the system parameters does not change for different
reaction terms, there are differences
in the front speed relaxation to its asymptotic value.  
In the case of FKPP nonlinearity
(i.e. ``pulled'' fronts), without advection (${\mathbf u} = 0$), it is
known~\cite{vansa} that the front velocity relaxes algebraically slow to its
asymptotic value. Therefore one can expect some numerical difficulties 
to find out the value of $V_f$, in particular for the slow reaction case
in which the front may interest a very large spatial region.
With this ``caveat'' in mind, in our simulations 
we varied the system size to carefully check
the convergence of $V_f$.
Using Arrhenius nonlinearity (i.e. ``pushed'' front) in the case 
without advection is known that the convergence to the asymptotic 
value is exponentially fast. Also in presence of a velocity field
we observe that the convergence is much faster than in the FKPP case. 

\section{Summary and conclusions} 
\label{sec:5} 

Enhancement of front propagation by an underlying flow is
a generic phenomenon for advection-reaction-diffusion systems.
A relevant question is how the front speed $V_f$ depends on the
detailed properties of the advecting velocity field,
in particular on the typical velocity $U$. 
For an arbitrary flow, it is extremely difficult to derive
this dependence analytically. 
Here, we have shown that for
all incompressible flows there exists an upper bound to the front
speed that links it to a single global property of the flow,
its effective diffusion coefficient. The analytic derivation 
of the effective diffusivity for a given velocity
field is itself a daunting task, but several generic properties 
are known and some exact results
are available for simple flows.
In the special case of fast advection, as compared to 
the reaction timescale $\tau_r$,
the upper bound is sharp, and therefore it is possible to obtain the 
dependence of $V_f$ on $U$. When molecular diffusivity is small
we have $V_f \sim U$ for flows with open streamlines such as the shear flow,  
and $V_f \sim U^{1/4}$ for cellular flows.
For slow advection, the bound ceases to be effective,
and one has to resort to numerical simulations in order to determine
the front speed. We find that for open-streamline flows there is still a
linear dependence $V_f \sim U$ whereas cellular flows display a 
$V_f \sim U^{3/4}$ dependence. 
Within the class of initial/boundary
conditions for which no flame quenching effect ever takes place,
those scaling laws appear to be universal with respect to the 
details of the reaction mechanism.\\ 
Which lessons can we draw from the present results 
for the open, challenging problem of front propagation in
turbulent flows? The main point that we want to
emphasize is the central role played by the effective
diffusion process in determining the front speed.
That is a reflection of the  
deep-rooted link between front propagation and transport properties.
The knowledge of turbulent transport
has experienced a significant progress in the past few years 
(see e.g. Refs \cite{SS00} and \cite{FGV}). 
We believe that those results will reveal 
helpful to shed light on the issue of front propagation in 
turbulent flow (for a similar point of view, see Ref.\cite{CY}).  


\section{Acknowledgments}
This work has been partially supported by INFM (PRA-TURBO), by the
European Network {\it Intermittency in Turbulent Systems} (contract
number FMRX-CT98-0175) and the MURST {\it cofinanziamento 1999} 
``Fisica statistica e teoria della materia condensata''. 
M.A. has been supported by the European Network 
{\it Intermittency in Turbulent Systems}.

\appendix
\section{Homogenization regime}
\label{app:a}
In this appendix we present briefly the application of 
homogenization techniques~\cite{MajdaBensou} to the 
reaction-advection-diffusion equation (\ref{eq:rad}).
We are interested in the large-time, large-scale asymptotics
for slow reaction timescales. Introducing the small parameter
$\epsilon$, we consider reaction times $\tau_r=\bar{\tau}_r\epsilon^{-2}$,
and look at the solutions of Eq.~(\ref{eq:rad}) for times $O(\epsilon^{-2})$
and scales $O(\epsilon^{-1})$. The separation in time-scales
allows a multi-scale treatment. 
Slow variables ${\mathbf X}=\epsilon{\mathbf x}$ and $T=\epsilon^2 t$
are introduced along with the fast variables ${\mathbf x}$ and $t$.
Slow and fast variables are considered as being independent.
As a consequence space-derivatives act as $\partial_x + \epsilon \partial_X$
and the time-derivative as $\partial_t + \epsilon^2 \partial_T$.
The concentration field
is expanded in a power series in $\epsilon$ as 
$\theta({\mathbf x},t,{\mathbf X},T)=
\theta^{(0)}({\mathbf x},t,{\mathbf X},T)+
\epsilon \theta^{(1)}({\mathbf x},t,{\mathbf X},T)+ \ldots$,
and this expression is plugged in Eq.~(\ref{eq:rad}).
At zero-th order in $\epsilon$ the equation reads
$$
\partial_t \theta^{(0)} + {\mathbf u}\cdot{\mathbf \nabla}_x \theta^{(0)}
= D_0 \nabla^2_x \theta^{(0)} \;. $$ 
Due to the dissipative nature 
of the latter equation, the solution at zero-th order will
decay to its average value on fast timescales:
$$
\theta^{(0)}({\mathbf x},t,{\mathbf X},T)=
\theta^{(0)}({\mathbf X},T)\,\,.
$$
At order $\epsilon$, we obtain the linear equation
$$
\partial_t \theta^{(1)} + {\mathbf u}\cdot{\mathbf \nabla}_x \theta^{(1)}
- D_0 \nabla^2_x \theta^{(1)}= 
- {\mathbf u}\cdot{\mathbf \nabla}_X \theta^{(0)}
$$ 
which allows the solution
$$
\theta^{(1)}({\mathbf x},t,{\mathbf X},T)=
\theta^{(1)}({\mathbf X},T)+{\mathbf w}({\mathbf x},t)
\cdot{\mathbf \nabla}_X \theta^{(0)}({\mathbf X},T)\;,
$$
provided that the auxiliary field ${\mathbf w}$ obeys the equation
$$
\partial_t{\mathbf w}  + {\mathbf u}\cdot{\mathbf \nabla}_x {\mathbf w}
= D_0 \nabla^2_x {\mathbf w}
- {\mathbf u}\;.
$$
Remark that the production term has not shown up yet.
It is at order $\epsilon^2$ that it enters the scene
$$
\partial_t \theta^{(2)} + {\mathbf u}\cdot{\mathbf \nabla}_x \theta^{(2)}
- D_0 \nabla^2_x \theta^{(2)}=-\partial_T \theta^{(0)} 
-  {\mathbf u}\cdot{\mathbf \nabla}_X \theta^{(1)} + 
D_0 \nabla^2_X \theta^{(0)} + 
$$
$$
2 D_0 {\mathbf \nabla}_x \cdot  {\mathbf \nabla}_X  \theta^{(1)}
+\frac{1}{\bar{\tau}_r} f(\theta^{(0)}) \;.
$$
The solvability condition for the equation at second order
requires that
$$
\partial_T \theta^{(0)}=D_0 \nabla^2_X \theta^{(0)}-
\langle {\mathbf u}\cdot{\mathbf \nabla}_X  \theta^{(1)} \rangle
+\frac{1}{\bar{\tau}_r} f(\theta^{(0)})
$$
where the brackets denote the average over the fast variables.
Plugging the expression for $\theta^{(1)}$ into the solvability 
condition yields the effective, large-scale equation
$$
\partial_T \theta^{(0)}= \sum_{i,j}D_{\mbox{\scriptsize eff}}^{\,\,ij} 
     \partial^2_{X^iX^j} \theta^{(0)}
+\frac{1}{\bar{\tau}_r} f(\theta^{0})
$$
where $D_{\mbox{\scriptsize eff}}$ is in general a tensor
with components
$D^{\,\,ij}_{\mbox{\scriptsize eff}} = D_0 \delta^{ij} -
{1 \over 2} \langle u^iw^j+u^jw^i \rangle 
$.
Considering the propagation in the $x$-direction only,
we recover an effective equation which is equivalent to
Eq.~(\ref{eq:rad}) with an effective diffusivity.
Therefore the front propagates at a maximal speed (computed in
the original, fast variables) given by 
$V_f \leq 2\sqrt{\frac{D_{\mbox{\scriptsize eff}}}{\tau_r}
\left(\sup_\theta \frac{f(\theta)}{\theta}\right) }$,
for FKPP nonlinearity one has 
$V_f = 2\sqrt{\frac{D_{\mbox{\scriptsize eff}}}{\tau_r}}$,
where $D_{\mbox{\scriptsize eff}} = D_{\mbox{\scriptsize eff}}^{\,\,11}$.

\section{Front speed for discrete time maps}
\label{app:b}
The front speed (\ref{eq:vfdiscrkpp}) for discrete maps
can be obtained by simple considerations just following
the standard way used for the derivation of $V_f$ in the
continuous time limit:
\be
   \frac{\partial}{\partial t} \theta =
   D_0 \nabla^2 \theta + \frac{1}{\tau_r} f(\theta)\,\,.
   \label{eq:rdapp}
\ee
Let us consider a front propagating from left to right.
For the sake of simplicity we discuss a FKPP nonlinearity
for the one dimensional case.
For $x \to \infty$, $\theta(x,t)$ has an exponential shape,
\be
   \theta(x,t)= e^{at - bx}\,\,,
   \label{eq:espshapeapp}
\ee
up to exponentially subleading terms.
Inserting (\ref{eq:espshapeapp}) in (\ref{eq:rdapp})
and linearizing around $\theta = 0$ one has:
\be
   a=D_0b^2 + {1 \over \tau_r} f'(0)\,\,.
   \label{eq:relabapp}
\ee
A saddle point argument gives a selection criterion
which allows for the determination of the front speed~\cite{vansa}:
\be
   V_f = \min_b {a(b) \over b} = 2 \sqrt{{D_0 \over \tau_r} f'(0)}\,\,.
   \label{eq:vf}
\ee
Consider now the discrete time reaction case, i.e.:
\be
   \frac{\partial}{\partial t} \theta = D_0 \nabla^2 \theta + 
   \sum_{n=-\infty}^\infty g(\theta) \delta(t - n)\,\,,
   \label{eq:raddiscr}
\ee
where for sake of simplicity we adopt $\Delta t=1$
(see Eq.~(\ref{eq:reacdimp})).\\
Indicating with $G(\theta)$ the reaction map 
(see Eq.~(\ref{eq:reactmap})) one has:
$$\theta(x,t+0) = G(\theta(x,t-0))\,\,.$$
Integrating the diffusion equation $\partial_t\theta=D_0\nabla^2\theta$
between $t+0$ and $t+1-0$ one has:
\be
   \theta(x, t+1 -0) = \frac{1}{\sqrt{2\pi}}
       \int e^{-\frac{w^2}{2}} G \left ( 
       \theta(x - \sqrt{2 D_0}\,w, t + 0) \right ) {\mathrm d} w\,\,,
   \label{eq:fwykacdimpmapapp}
\ee
Assuming the shape (\ref{eq:espshapeapp}) and linearizing around
$\theta = 0$, i.e.: $G(\theta) \simeq G'(0) \theta$, a simple
gaussian integration gives:
$$ e^{a(t+1)-bx} \sim e^{\ln G'(0) + D_0b^2-bx+at}\,\,.$$
The above result implies
$$ a= \ln G'(0) + D_0b^2\,\,$$
this is nothing but Eq.~(\ref{eq:relabapp}) now with $\ln G'(0)$
instead of ${1 \over \tau_r} f'(0)$. The same selection criterion
gives
\be
   V_f = 2 \sqrt{D_0\ln G'(0)}\,\,.
   \label{eq:vfdiscr}
\ee

\section{Numerical method}
\label{app:c}
\begin{center}
\begin{figure}[htb]
\epsfig{figure=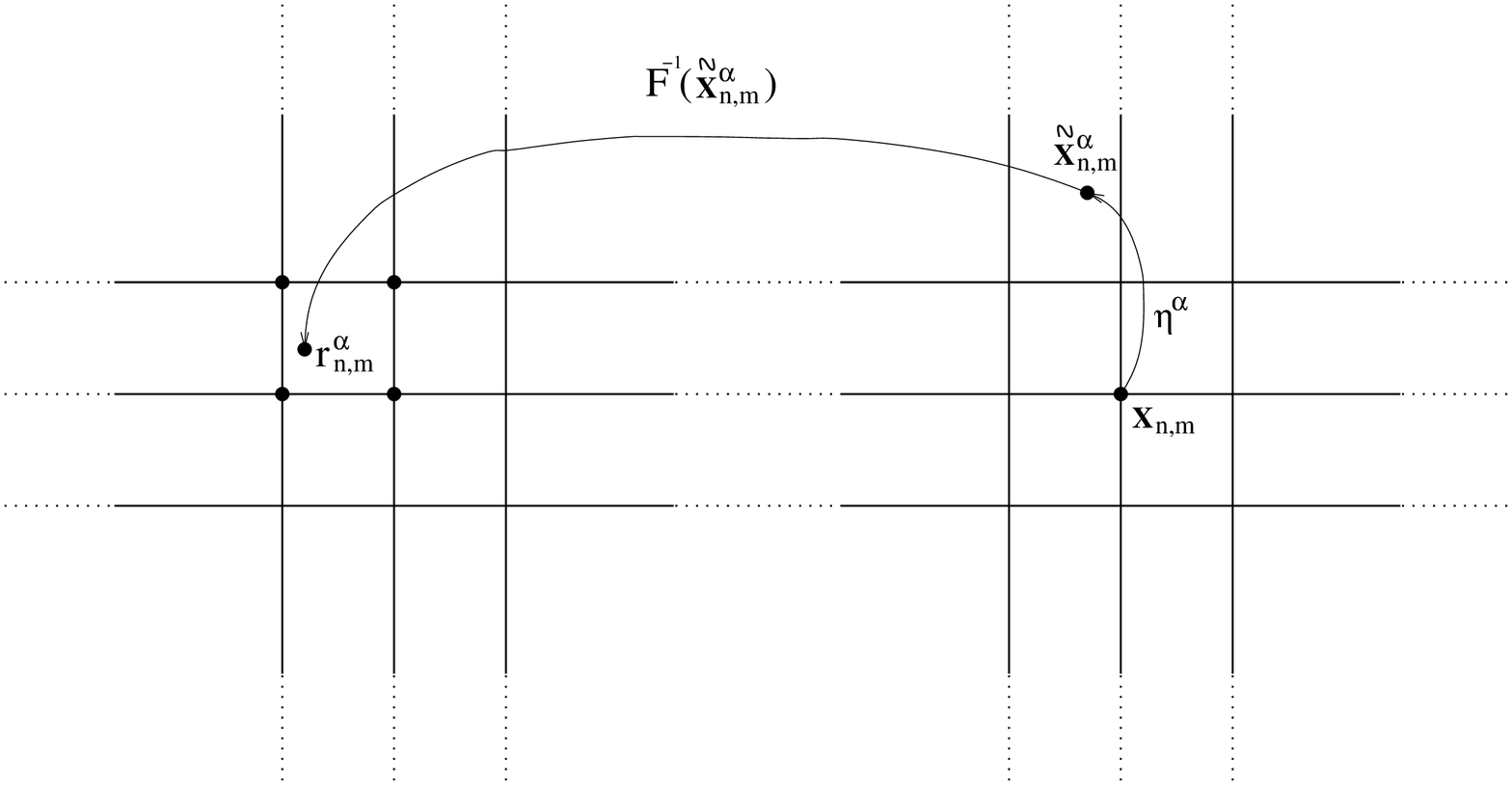,width=0.95\textwidth,angle=0}\hfill
\caption{Pictorial scheme of the numerical algorithm,
as discussed in Appendix C. Here,
$\eta^\alpha = \sqrt{2D_0\,\Delta t}\, {\mathbf W}^\alpha$ 
where ${\mathbf W}^\alpha$ is a standard gaussian variable.}
\label{fig:nummet}
\end{figure}
\end{center}

Since we are interested in propagation along the $x$-axis,
we consider a slab with sides $L_x \gg L_y$. The boundary
conditions are periodic in the $y$-direction
$\theta(x,y,t)=\theta(x,y+L_y,t)$. To fulfill the conditions (\ref{bc-x})
$\lim_{x \to \infty} \theta(x,y,t)=0$ and
$\lim_{x \to -\infty} \theta(x,y,t)=1$ numerically, we set
$\theta(0,y,t)=1$ and $\theta(L_x,y,t)=0$, which is a good
approximation as long as the fronts leading edge has not reached $L_x$.
We introduce a lattice of step size $\Delta x$ and $\Delta y$ (for sake of
simplicity we assume $\Delta x = \Delta y$) so that the field
$\theta(x,y,t)$ is defined on the points 
${\mathbf x}_{n,m} = (n\Delta x,m\Delta y)$.
The numerical code computes $\theta_{n,m}(t+\Delta t) = \theta(n\Delta x,
m\Delta y, t+\Delta t)$ in terms of $\theta_{n,m}(t)$ using 
Eq.~(\ref{eq:feykacmap}). For each grid point ${\mathbf x}_{n,m}$, 
one introduces $N$ independent standard gaussian
variables ${\mathbf W}^\alpha$, $\alpha=1,\dots,N$, $N\gg 1$, and computes
${\mathbf {\tilde x}}^\alpha_{n,m}={\mathbf x}_{n,m} - 
  \sqrt{2D_0\,\Delta t}\, {\mathbf W}^\alpha$ and finally from this 
${\mathbf r}^\alpha_{n,m} = {\mathbf F}^{-1}({\mathbf \tilde{x}}_{n,m}^\alpha)$. 
For $\theta_{n,m}(t+\Delta t)$ one needs the values of $\theta$ at
time $t$ in the positions ${\mathbf r}_{n,m}^\alpha$. In general the
${\mathbf r}_{m,n}^\alpha$ are not on the grid points $(n \Delta x, m \Delta y)$, 
nevertheless we can compute the value $\theta({\mathbf r}_{n,m}^\alpha,t)$
using linear interpolation from $\theta_{n,m}(t)$.
Therefore we have
\begin{equation}
\theta_{n,m}(t+\Delta t) = {1\over N} \sum_{\alpha=1}^N 
   G[\theta({\mathbf r}_{n,m}^\alpha,t)].
\end{equation} 
Typically one has a good convergence for $N=50$.
To simulate the diffusion process we have to impose a relation between 
$D_0$, $\Delta x$ and $\Delta t$ to insure that the diffusion transports
a particle over distances larger than the grid-size,
${\sqrt{2D_0\,\Delta t}\over \Delta x} > 1$ (see Fig~\ref{fig:nummet}).

 

\newpage

\newpage
\newpage
\newpage
\newpage


\newpage
\begin{thebibliography}{99} 
  
\bibitem{flame} F.~A.~Williams, {\em Combustion Theory\/}, 
Benjamin-Cummings, Menlo Park (1985);
E.~R.~Abraham, Nature {\bf 391}, 577 (1998);
I.~R.~Epstein, Nature {\bf 374}, 231 (1998);
J.~Ross, S.~C.~M\"uller, and C.~Vidal, Science {\bf 240}, 
460 (1988).

\bibitem{combus} N.~Peters, {\em Turbulent Combustion\/}, 
Cambridge Univ. Press, Cambridge UK (2000).

\bibitem{Malham} S.~Malham and J.~Xin, 
Comm. Math. Phys {\bf 199}, 287 (1998).

\bibitem{kpp} A.~N.~Kolmogorov, I.~G.~Petrovskii, and N.~S.~Piskunov, 
Moscow Univ. Bull. Math. {\bf 1}, 1 (1937);
R.~A.~Fischer, Ann. Eugenics {\bf 7}, 355 (1937). 

\bibitem{xin2000} J.~Xin, SIAM Review {\bf 42}, 161 (2000).

\bibitem{quenching} P.~Constantin, A.~Kiselev and L.~Ryzhik, 
LANL e-print archive nlin/0006024 (2000).

\bibitem{AronWein78} D.~G.~Aronson and H.~F.~Weinberg,
Adv. Math. {\bf 30}, 33 (1978).

\bibitem{Armero97} J.~Armero, A.~Lacasta, L.~Ramirez-Piscina, J.~Casademunt, J.~M.~Sancho and F.~Sagues, Phys Rev E {\bf 56}, 5405 (1997).

\bibitem{vansa} U.~Ebert and W.~Van Saarloos, 
Physica D {\bf 146}, 1 (2000). 

\bibitem{CW} P.~Clavin, and F.~A.~Williams,
J. Fluid Mech. {\bf 90}, 589 (1979). 

\bibitem{KA} A.~R.~Kerstein, and W.~T.~Ashurst,
Phys. Rev. Lett {\bf 68}, 934 (1992).

\bibitem{Ros} M.~N.~Rosenbluth, A.~L.~Berk, I.~Doxas, and W.~Horton, 
Phys. Fluids {\bf 30}, 2636 (1987).

\bibitem{Shr} B.~I.~Shraiman, 
Phys. Rev. A {\bf 36}, 261 (1987).

\bibitem{Pom} Y.~Pomeau
C. R. Acad. Sci. {\bf 301}, 1323 (1985).

\bibitem{Rhi} P.~B.~Rhines and W.~R.~Young, 
J. Fluid Mech. {\bf 133}, 133 (1983).

\bibitem{Ronney95} P.~D.~Ronney, in {\it Modeling in Combustion
Science} 3, Eds. J.~Buckmaster and T.~Takeno
(Springer, New York 1995).

\bibitem{Const} P.~Constantin, A.~Kiselev, A.~Oberman, and L.~Ryzhik,
LANL e-print archive math/9907132 (1999).

\bibitem{KR00} A.~Kiselev and L.~Ryzhik, 
LANL e-print archive math/0002175 (2000).

\bibitem{mclaughlin} R.~M.~McLaughlin and J.~Zhu,
Comb. Sci. and Tech. {\bf 129}, 89 (1997).

\bibitem{embid} P.~F.~Embid, A.~Majda and P.~E.~Sounganidis,
Phys. Fluids {\bf 7}, 2052 (1995).

\bibitem{yakhot} U.~Yakhot, Comb. Sci. and Tech. {\bf 60}, 191 (1988).

\bibitem{pocheau} A.~Pocheau, Phys. Rev. E {\bf 49}, 1109 (1994).

\bibitem{marti} A.~C.~Mart\'{\i}, F.~Sagues and J.~M.~Sancho,
Phys. Fluids {\bf 9}, 3851 (1997).

\bibitem{aref} H.~Aref, J. Fluid Mech. {\bf 143}, 1 (1984);
J.~M.~Ottino, Ann. Rev. Fluid Mech. {\bf 22}, 207 (1990);
A.~Crisanti, M.~Falcioni, G.~Paladin and A.~Vulpiani,
La Rivista del Nuovo Cimento {\bf 14} (12), 1 (1991). 

\bibitem{freid} M.~Freidlin, {\em Markov Processes and 
Differential Equations\/} Birkhauser, Boston (1996).

\bibitem{fedotov}S.~Fedotov, Phys. Rev. E {\bf 55} 2750 (1997).

\bibitem{AM} M.~Avellaneda, and A.~Majda, 
Phys. Rev. Lett. {\bf 62}, 753 (1989).          

\bibitem{Castiglione} P.~Castiglione, A.~Crisanti, A.~Mazzino,
M.~Vergassola and A.~Vulpiani, J. Phys. {\bf A 31}, 7197 (1998).

\bibitem{Zaslavsky} G.~M.~Zaslavsky, D.~Stevens and H.~Weitzner,
Phys. Rev. E {\bf 48}, 1683 (1993).

\bibitem{Zeldovich} Ya.~B.~Zel'dovich, 
Sov. Phys. Dokl. {\bf 27}, 797 (1982).

\bibitem{vladimirova} N.~Vladimirova, F.~Cattaneo, A.~Malagoli,
A.~Oberdam and O.~Ruchayskiy, preprint 2000.

\bibitem{AV} M.~Avellaneda and M.~Vergassola
Phys. Rev. E {\bf 52}, 3249 (1995).

\bibitem{K70} R.~H.~Kraichnan, in {\em The Pad\'e Approximants in 
Theoretical Physics\/}, Acad. Press Inc., New York (1970).

\bibitem{Taylor21} G.~I.~Taylor, Proc. London Math. Soc. {\bf 20},
196 (1921).

\bibitem{MK99} A.~J.~Majda, and P.~R.~Kramer, Phys. Rep. {\bf 314}, 237
 (1999).
\bibitem{Pomeau} B.~Audoly, H.~Beresytcki and Y.~Pomeau
C. R. Acad. Sci. {\bf 328}, S\'erie II b, 255 (2000).

\bibitem{ronney} P.~D.~Ronney, B.~D.~Haslam and N.~O.~Rhys, 
Phys. Rev. Lett. {\bf 74}, 3804 (1995). 

\bibitem{LichLieb} A.~J.~Lichtenberg and M.~A.~Liebermann,
{\it Regular and Chaotic Dynamics} (Springer-Verlag, Berlin 1992).


\bibitem{SS00} B.~Shraiman and E.~Siggia, Nature {\bf 405}, 639 (2000).

\bibitem{FGV} G.~Falkovich, K.~Gaw\c{e}dzki and M.~Vergassola,
Rev. Mod. Phys., in press. (2001).

\bibitem{CY} M.~Chertkov and V.~Yakhot, Phys. Rev. Lett.
{\bf 80 }, 2837 (1998).

\bibitem{MajdaBensou} A.~J.~Majda and P.~R.~Kramer, 
Phys. Rep. {\bf 314}, 237 (1999); 
A.~Bensoussan, J.~Lions and G.~Papanicolaou
{\it Asymptotic Analysis for Periodic Structures}
(North Holland, Amsterdam 1978).

\end{thebibliography}
\end{document}